\documentclass{emulateapj}
\usepackage{graphicx}
\usepackage{amsmath}
\usepackage{natbib}
\submitted{The Astrophysical Journal, 604:632-652, 2004 April 1}
\newcommand{\tbfrac}[2]{\genfrac{}{}{0pt}{0}{#1\strut}{#2\strut}}
\begin{document}
\title{Formation of Massive Black Holes in Dense Star Clusters. I. 
Mass Segregation and Core Collapse}

\author{M. Atakan G\"urkan\altaffilmark{1}, 
Marc Freitag\altaffilmark{2}, and
Frederic A.\ Rasio\altaffilmark{1}}
\altaffiltext{1}{Dearborn Observatory, Department of Physics and Astronomy, Northwestern University, 2131 Tech Drive, Evanston, IL 60208, USA;
{\tt ato@northwestern.edu}, {\tt rasio@northwestern.edu}}
\altaffiltext{2}{Astronomisches Rechen-Institut, M\"onchhofstrasse 12-14,\\
D-69120 Heidelberg, Germany; {\tt freitag@ari.uni-heidelberg.de}}
\begin{abstract}
 We study the early dynamical evolution of young, dense star clusters
 using Monte Carlo simulations for systems with up to $N = 10^7$
 stars. Rapid mass segregation of massive main-sequence stars and the
 development of the Spitzer instability can drive these systems to
 core collapse in a small fraction of the initial half-mass relaxation
 time. If the core collapse time is less than the lifetime of the
 massive stars, all stars in the collapsing core may then
 undergo a runaway collision process leading to the formation of a
 massive black hole. Here we study in detail the first step in this
 process, up to the occurrence of core collapse. We have performed
 about 100 simulations for clusters with a wide variety of initial
 conditions, varying systematically the cluster density profile,
 stellar IMF, and number of stars. We also considered the effects of initial 
 mass segregation and stellar evolution mass loss. Our results show that, for
 clusters with a moderate initial central concentration and any realistic
 IMF, the ratio of core collapse time to initial half-mass relaxation
 time is typically $\sim 0.1$, in agreement with the value previously found
 by direct $N$-body simulations for much smaller systems. Models with even higher 
 central concentration initially, or with initial mass segregation (from star formation) 
 have even shorter core-collapse times. Remarkably, we find that, for all
 realistic initial conditions, the mass of the collapsing core is
 always close to $\sim10^{-3}$ of the total cluster mass, very similar
 to the observed correlation between central black hole mass and total
 cluster mass in a variety of environments. We discuss the implications of our results
 for the formation of intermediate-mass black holes in globular clusters and super star 
 clusters, ultraluminous X-ray sources, and seed black holes in proto-galactic
 nuclei.
\end{abstract}

\section{Introduction}

\subsection{Astrophysical Motivation}

It is now well established that the centers of most galaxies host
supermassive black holes (BH) with masses in the range
$M_{\textstyle\bullet} \sim 10^6-10^9\,M_\odot$
\citep{KG01,FerrareseEtAl01}. The evidence is particularly compelling for a BH
of mass $\simeq4\times10^6\,M_\odot$ at the center of our own Galaxy
\citep{GhezEtAl00,EckartEtAl02,SchodelEtAl02,GhezEtAl03}.  Dynamical
estimates indicate that, across a wide range, the central BH mass is
about 0.1\% of the spheroidal component of the host galaxy
\citep{Ho98}. A related correlation may exist with the total
gravitational mass of the host galaxy (basically the mass of its dark
matter halo) \cite{Ferrarese02}.  An even tighter correlation is
observed between the central velocity dispersion and the central BH
mass \citep{FM00,GebhardtEtAl00,Tetal02}.

Theoretical arguments and recent observations suggest
that a central BH may also exist in some globular clusters 
\citep{vdM01,vdM03}. 
In particular, recent HST
observations and dynamical modeling of M15 by \cite{Getal02,Getal03} yielded results
that are consistent with the presence of a central massive BH in
this cluster. Similarly \citet{GRH02} have argued for the
existence of an even more massive BH at the center of the
globular cluster G1 in M31. However, $N$-body simulations
\citep{Betal03,Betal03b} suggest that the observations of
M15 and G1 could be explained equally well by the presence of many compact
objects near the center without a massive BH \citep{vdM01}.

When the correlation between the mass of the central BH and the spheroidal
component in galaxies is extrapolated to smaller stellar systems like globular 
clusters, the inferred BH
masses are $\sim 10^3-10^4\,M_\odot$, much larger than a $\sim10\,M_\odot$ stellar-mass 
BH,  but much
smaller than the $\sim10^6-10^9\,M_\odot$ of supermassive BH. Hence, these are
called {\em intermediate-mass black holes\/} (IMBH). If some globular clusters
do host a central IMBH the question arises of how these objects were formed
\citep[for recent reviews see][]{vdM03,RFG03}.
One natural path for their formation in any young stellar system with
a high enough density is through runaway collisions and mergers of
massive stars following core collapse.  These runaways could easily
occur in a variety of observed young star clusters such as the ``young
populous clusters'' like the Arches and Quintuplet clusters in our
Galactic center and the ``super star clusters'' observed in all
starburst and galactic merger environments \citep[see,
e.g.,][]{FKMSRM99,GS99}. The Pistol Star in the Quintuplet cluster
\citep{FigerEtAl98} may be the product of such a runaway, as
demonstrated recently by direct $N$-body simulations \citep{PZM02}. A
similar process may be responsible for the formation of seed BH in
proto-galactic nuclei, which could then grow by gas accretion or by
merging with other IMBH formed in young star clusters
\citep{Ebisuzakietal01,HM03}.  Further observational evidence for IMBH
in dense star clusters comes from recent {\em Chandra\/} and
XMM-{\em Newton\/} observations of ultra-luminous X-ray sources, which are
often (although not always) associated with young star clusters and
whose high X-ray luminosities in many cases suggest a compact object
mass of at least $\sim 10^2\, M_\odot$
\citep{Kaaretetal01,Ebisuzakietal01,MFMF03},
although beamed emission by an accreting stellar-mass BH provides an
alternative explanation \citep{Kingetal01,ZF02}.

\begin{figure*}
\begin{center}
\resizebox{0.9\hsize}{!}{\includegraphics[clip]{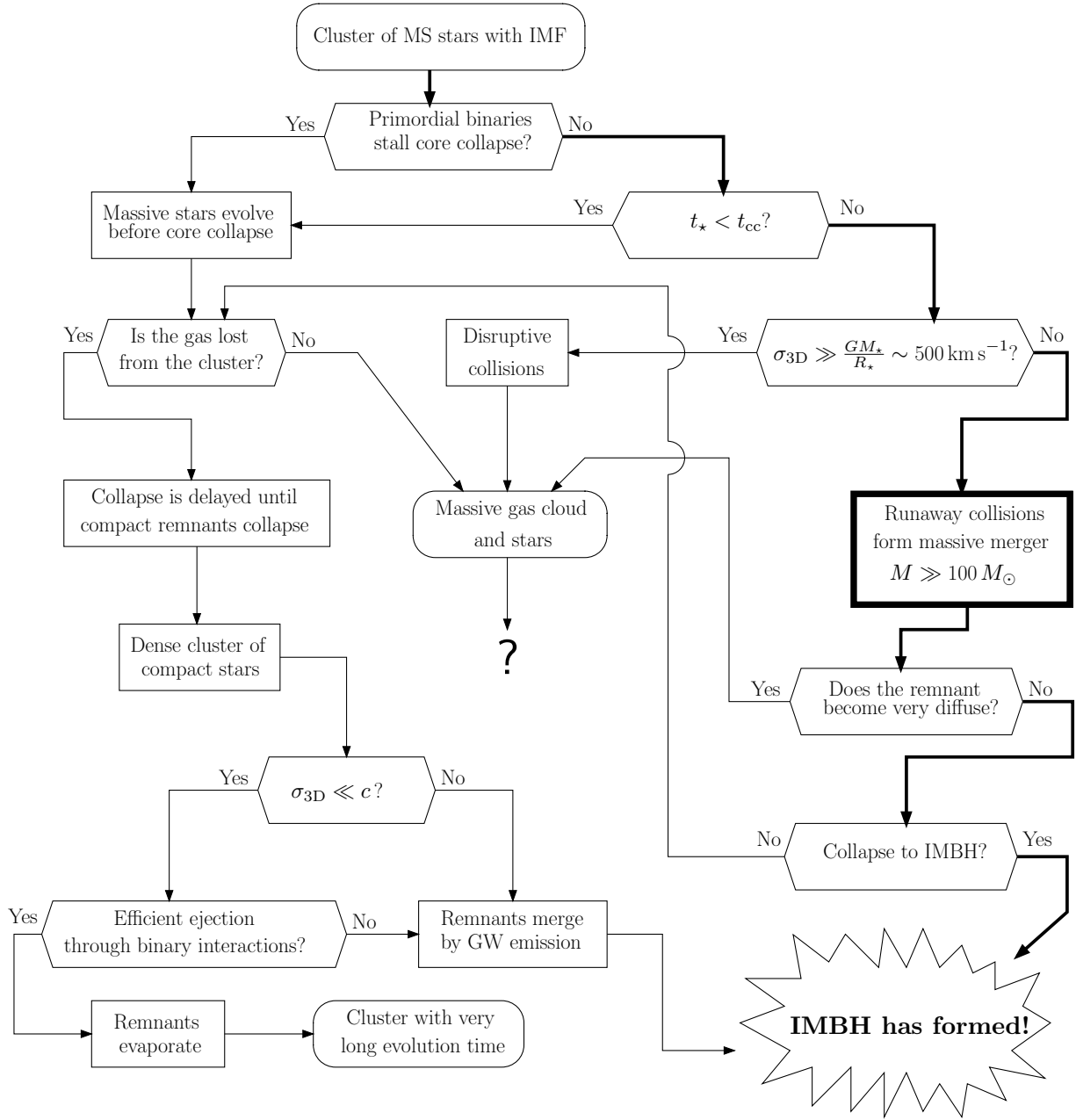}}
\caption{Various possible scenarios for the early dynamical evolution of a dense star cluster 
with a realistic IMF. The path for the formation of an
IMBH through core collapse and runaway collisions (studied in this paper) is indicated by a thicker line. 
An alternative scenario involves successive mergers of stellar-mass BH binaries driven by
a combination of dynamical interactions and gravitational radiation (left side
of the diagram). For a runaway to occur, the core collapse time $t_{\rm cc}$ must be
smaller than the stellar lifetime $t_*$ of the most massive stars in the cluster.
High-velocity disruptive collisions, the formation of a very extended and diffuse merger
remnant, or the accumulation in the cluster of gas released by stellar winds and supernova 
explosions could lead to the formation of a complex system containing stars embedded in a 
dense gas clouds. The final fate of such a system is highly uncertain.
 \label{fig_rees_diag}
}
\end{center}
\end{figure*}

When they are born, star clusters are expected to contain many young stars
with a wide range of masses, from $\sim0.1 M_\odot$ to $\sim100 M_\odot$,
distributed according to a Salpeter-like initial mass function (IMF) \citep{Clarke03}. 
Inspired by an early version from \cite{Rees84} for the formation of supermassive BH, 
we show in Figure~\ref{fig_rees_diag} a diagram illustrating the two
main scenarios leading to the formation of an IMBH at the center of a dense star 
cluster. The early
stages of dynamical evolution are dominated by the stars in the upper part
of the mass spectrum. Through dynamical friction, these heavy stars tend to
concentrate toward the center and drive the system to core collapse.
Successive collisions and mergers of the massive stars during core collapse
can then lead to a runaway process and the rapid formation of a very massive
object containing the entire mass of the collapsing cluster core. Although
the fate of such a massive merger remnant is rather uncertain, direct
``monolithic'' collapse to a BH with no or little mass loss is a likely
outcome, at least for sufficiently small metallicities \citep{HFWLH02}. An
essential condition for this runaway to occur is that the core collapse
must occur before the most massive stars born in the cluster end
their lives in supernova explosions \citep{PZM02,RFG03}.

The accumulation at the center of a galaxy of many IMBH produced 
through this runaway process in nearby young star clusters (like the Arches and Quintuplet clusters
in our Galaxy) provides an interesting new way of building up
the mass of a central supermassive BH \citep{PZM02}.
It is possible that this process of accumulation is still ongoing in our own
Galaxy \citep{HM03}.
These ideas have potentially important implications for
the study of supermassive BH by the Laser Interferometer Space Antenna (LISA), since the 
inspiral of IMBH
into a supermassive BH provides the best source of low-frequency 
gravitational waves for direct study of strong field gravity \citep{CT02}. 

In the alternative scenario where massive stars evolve and produce supernovae
before the cluster goes into core collapse, a subsystem of
stellar-mass BH will be formed (Figure~\ref{fig_rees_diag}).
As demonstrated in
Section~\ref{subsec_stell_evol}, the mass loss from supernovae provides
significant indirect heating of the cluster core, delaying the onset
of core collapse until much later, after the stellar remnants undergo
mass segregation. The final fate of a cluster with a component of
stellar-mass BH remains highly uncertain. This is because realistic
dynamical simulations for such clusters (containing a large number of
black holes {\em and\/} ordinary stars with a realistic mass spectrum) 
have yet to be performed. 
For old and relatively small systems (such as Galactic
globular clusters), complete evaporation is likely (with all the
stellar-mass BH ejected from the cluster through 3-body and 4-body
interactions in the dense core). This is expected theoretically on the
basis of simple qualitative arguments
\citep{KHMcM93,SH93} and has been demonstrated recently by direct
$N$-body simulations for very small systems containing only 
$\sim 10$ BH \citep{PZMcM00}. 
However, for larger systems (more massive globular clusters
or proto-galactic nuclei), contraction of the cluster to a highly
relativistic state could again lead to successive mergers (driven by
gravitational radiation) and the formation of a single massive BH
\citep{QS89,Lee95,Lee01}. Moreover, it has been recently 
suggested that, if stellar-mass BH are formed with a broad mass spectrum (a
likely outcome for stars of very low metallicity; see \citealt{HWFL03}), the
most massive BH could resist ejection, even in a system with low escape 
velocity such as a globular cluster. These more massive BH could then grow
by repeatedly forming binaries (through exchange interactions)
with other BH and merging with their companions \citep{MH02}. However, as 
most interactions will probably result in the ejection 
of one of the lighter BH, it is unclear whether any object could grow 
substantially through this mechanism before running out of companions to merge
with. 

\subsection{Core Collapse and the Spitzer Instability}
\label{CC_and_SI}
The physics of gravothermal contraction and core collapse is by now very well 
understood for {\it single-component\/} systems (containing all equal-mass stars).
In particular, the dynamical evolution of an isolated, single-component Plummer sphere
to core collapse has been studied extensively and
has become a testbed for all numerical codes used to compute the 
evolution of dense star clusters
\citep{AHW74,GS94}. This evolution can be
visualized easily using Lagrange radii, enclosing a fixed
fraction of the total mass of the system \citetext{see, e.g.,
Fig.~3 of \citealp{JRPZ00} and Fig.~5 of \citealp{FB01}}. 
As the system evolves, the inner Lagrange radii contract while the outer
ones expand. This so-called gravothermal contraction
results from the negative heat capacity that is a common property of all
gravitationally bound systems \citep{EHI87}. In the absence of an energy
source in the cluster
core, and in the point-mass limit (i.e., neglecting physical collisions between stars), 
the contraction of the cluster core becomes self-similar and continues
indefinitely. This phenomenon is known as core collapse and its universality is
very well established \citep[see, e.g.,][Table~1 and references
therein]{FB01}.

\begin{figure}
\resizebox{\hsize}{!}{\includegraphics[clip]{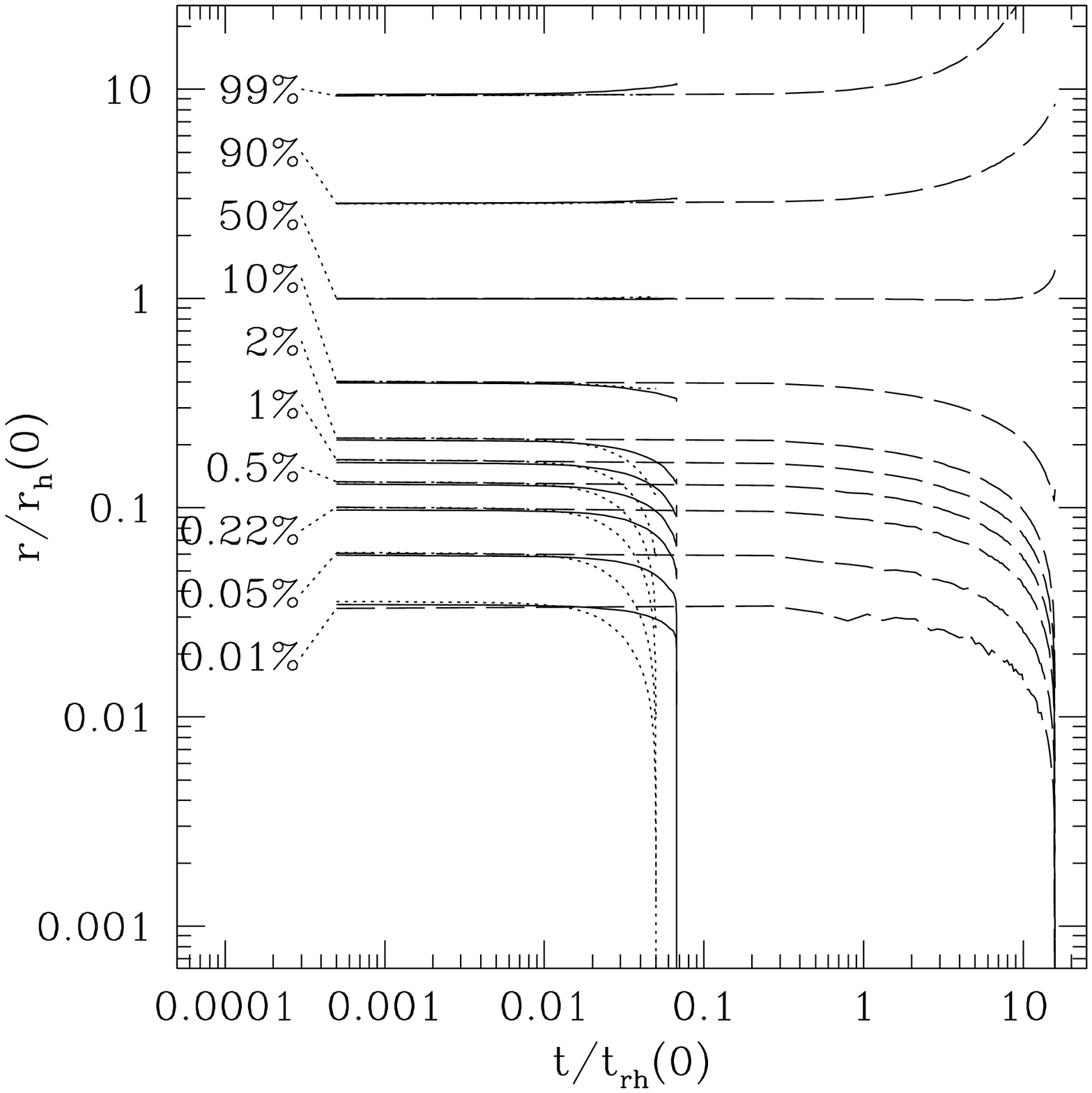}}
\caption{
  Evolution of single-component and Salpeter-IMF Plummer models to core collapse. 
  Lagrange radii enclosing 
  a constant mass fraction, indicated on the left, are shown as a function of time. The radii are
  in units of the initial half-mass radius of the cluster and time is in units of the initial 
  half-mass relaxation time.
  The solid lines are from our Monte Carlo simulation of a Plummer model containing 
  $1.25\times10^6$ stars with a
  Salpeter IMF (within $m_{\rm min}=0.2\,M_\odot$
  and $m_{\rm max}=120\,M_\odot$). The dotted lines show the result of a gaseous
  model simulation of the same cluster, with 50 discrete mass components approximating a
  Salpeter IMF (within the same mass limits). See Sec.~2 for more discussions of the various methods.
  For comparison, the dashed lines show the evolution 
  of a single-component Plummer Model (computed with our Monte Carlo code). Note our key result:
  the core collapse time is more than two orders of magnitude shorter for a cluster with
  a realistic IMF.
\label{fig_cc_compare}
}
\end{figure}

When the system contains stars with a
variety of masses, the evolution to core collapse is accelerated. This has
been demonstrated by the earliest $N$-body and Monte Carlo simulations 
\citep{Aarseth66,Henon71,Wielen75}.
To illustrate this behavior, and as a preview of the results presented later in Sec.~4,
we show in Figure~\ref{fig_cc_compare} the evolution of a cluster described initially by a simple
Plummer model containing $1.25\times10^6$ stars with a
broad Salpeter IMF between $m_{\rm min}=0.2\,M_\odot$
and $m_{\rm max}=120\,M_\odot$. Core collapse occurs after $\lesssim 0.1\,t_{\rm rh}(0)$, where
$t_{\rm rh}(0)$ is the initial half-mass relaxation time (see 
eq.~(\ref{eq_rel_time}) below). In sharp contrast,
core collapse in a single-component Plummer model occurs 
after $\ga 10\,t_{\rm rh}(0)$
(a well-known result). Thus the presence of a broad IMF can dramatically
accelerate the evolution of the cluster to core collapse. 

This acceleration of the evolution to core collapse is due to the
changing nature of energy transfer in the presence of a wide mass
spectrum. Relaxation processes tend to establish energy equipartition
\citep[see, e.g.,][Sec.~8.4]{BT87}. In a cluster where the masses
of the stars are nearly equal, this can be (very nearly) achieved.
The core collapse is then a result of energy transfer from the inner
to the outer parts of the cluster, leading to gravothermal
contraction \citep{LBW68,Larson70}.  A large difference between the
masses of the stars allows a more efficient mechanism for energy
transfer. In this case, energy equipartition would tend to bring the
heavier stars to lower speeds.  However, as a result, the heavier
stars sink to the center, where they tend to gain kinetic energy,
while the lighter stars move to the outer halo. This process is
called ``mass segregation.''  As the mass segregation proceeds, the
core contracts and gets denser, leading to a shorter relaxation time,
which in turn increases the rate of energy transfer from heavier to
lighter stars. In typical cases this evolution eventually makes the
heavier stars evolve {\it away\/} from equipartition
\citep{Spitzer69}.

The fundamental inability of the heavier stars to establish energy equipartition with the lighter
stars in a system with a continuous mass spectrum is similar to the Spitzer ``mass-segregation
instability'' in two-component clusters. 
\cite{Spitzer69},
using analytic methods and a number of simplifying assumptions, determined a simple criterion for
a two-component system to achieve energy equipartition in equilibrium. If the mass of the lighter
(heavier) stars is  $m_1$ ($m_2$) and the total mass in the light (heavy)
component is $M_1$ ($M_2$), then Spitzer's criterion can be written
\begin{equation}
\label{eqn_Spitzer}
S \equiv \left(\frac{M_2}{M_1}\right)\left(\frac{m_2}{m_1}\right)^{3/2} <
0.16.
\end{equation}
\citet{WJR00}, using numerical simulations, obtained a more accurate empirical
condition,
\begin{equation}
\label{eqn_Watters}
\Lambda \equiv 
\left(\frac{M_2}{M_1}\right)\left(\frac{m_2}{m_1}\right)^{2.4} < 0.32.
\end{equation} 
When this stability criterion is not satisfied, energy
equipartition cannot be established between heavy and light stars.
\citet{Spitzer69} noted that the
equilibrium would not be achieved for
realistic mass spectra, because there is always
sufficient mass in high-mass stars that, through mass segregation, they can
form a subsystem (near the center of the cluster) 
that decouples dynamically from the lower-mass stars. This was later supported by more
detailed theoretical studies \citep{SDY71,Vishniac78,IS85}.

\begin{figure}
\resizebox{\hsize}{!}{\includegraphics[clip]{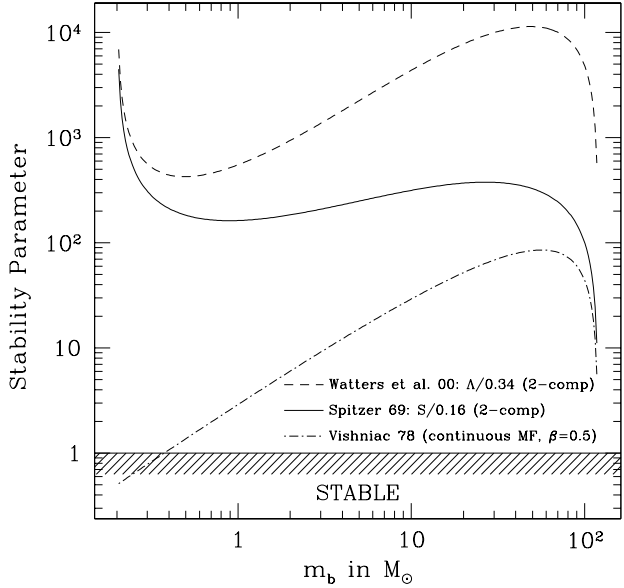}}
\caption{
Various criteria for the realization of equipartition in a cluster with
a Salpeter IMF. According to these, energy equipartition in dynamical
equilibrium can only be achieved if the corresponding curve lies
entirely under the dotted horizontal line. Conditions by Spitzer and
Watters et al.\ were devised for two-component models, so we divide
the mass function into stars lighter and heavier than some arbitrary
boundary value $m_{\rm b}$ and we evaluate the stability conditions
for any value of $m_{\rm b}$. Vishniac's analysis is genuinely adapted
to a continuous mass function. He derives a necessary condition for
stable equipartition that must be obeyed for any stellar mass $m_{\rm
b}$ in the range covered by the mass spectrum. See text for details.
\label{fig_ins}
}
\end{figure}

Here we carry out a simple calculation to show that a typical
system, with a Salpeter IMF (see Sec.~3.2) extending from $0.2\,M_\odot$ to
$120\,M_\odot$, when viewed as a two-component cluster of ``light
stars'' and ``heavy stars,'' does not satisfy any of the simple
stability criteria. Let us separate the stars into
these two components according to some arbitrary boundary: we
group all stars lighter than $m_{\rm b}$ in the first component,
and all stars heavier than $m_{\rm b}$ in the second component.  We use $m_1$
and $M_1$ to denote the values of the average and total mass in the first
component, and $m_2$ and $M_2$ similarly for the second component.
We obtain, with $m_{\rm b}$ in solar mass,
\begin{align}
\frac{M_1}{M_2}& = 
\frac{0.2^{-0.35}-m_{\rm b}^{-0.35}}{m_{\rm b}^{-0.35}-120^{-0.35}}, {\rm ~~~~~~~and}
\label{eqn_mb1}\\
\frac{m_1}{m_2}& =
\frac{M_1}{M_2}\cdot\frac{N_2}{N_1} =
\frac{0.2^{-0.35}-m_{\rm b}^{-0.35}}{m_{\rm b}^{-0.35}-120^{-0.35}}\cdot
\frac{m_{\rm b}^{-1.35}-120^{-1.35}}{0.2^{-1.35}-m_{\rm b}^{-1.35}}.
\label{eqn_mb2}
\end{align}
When these values are used in equations (\ref{eqn_Spitzer}) and
(\ref{eqn_Watters}) we see (Figure~\ref{fig_ins}) that
the stability criteria are almost never satisfied for
any value of $m_{\rm b}$, except when it is extremely close to the maximum 
mass (as expected, since the model reduces artificially to a single-component system in this 
limit, with all stars having the average mass).

\citet{Vishniac78}, devised a criterion genuinely 
adapted to clusters with a continuous mass spectrum, under the {\it
ad-hoc\/} assumption that the shape of the density distribution does not
depend on the stellar mass. He derives the following necessary condition for stability: 
\begin{equation}
	\beta\left(\frac{m_{\rm b}}{m_1}\right)^{3/2}\frac{M_2}{M_1 + M_2} < 1,\ \
	\mbox{with}\ \ \beta\simeq 0.5,
\end{equation} 
for all $m_{\rm b}$ (his eq.~17). Figure~\ref{fig_ins} shows that
this condition cannot be satisfied either\footnote{The applicability of
Vishniac's criterion appears somewhat questionable. Through FP simulations,
\citet{IS85} have shown that an IMF exponent $\alpha \gtrsim 6.0$ (see 
Sec.~\ref{subsec_mass_fcts}) is required for central equipartition to
set in before core collapse while Vishniac's criterion predicts that
$\alpha \gtrsim 3.5$ is sufficient.}.  These results suggest that for
any Salpeter-like IMF, one can always find a collection of stars that
have large enough average and total mass that they will dynamically
decouple from the lighter stars in the cluster. Once these decouple
they form a subsystem with a much shorter relaxation time and
consequently evolve to core collapse very rapidly.

The half-mass relaxation time (relaxation time at the half-mass radius $r_h$)
for a cluster of $N$ stars is given by \citep[][eq.~2.63]{Spitzer87}:
\begin{equation}
\label{eq_rel_time}
t_{\rm rh}(0) = \frac{0.138 N}{\ln(\gamma_{\rm c} N)}
\left(\frac{r_{\rm h}^3}{GM}\right)^{1/2}
\propto \frac{N/\ln(\gamma_{\rm c} N)}{\rho^{1/2}} \:,
\end{equation}
where $\rho$ is the mass density and $\gamma_{\rm c}\sim0.01$ 
in the Coulomb logarithm.
Let us assume that a collapsing subsystem is formed by stars that constitute
1\% of the total mass, and that all of them come from the uppermost part of the mass
spectrum. For a Salpeter IMF with $m_{\rm min}=0.2\,M_\odot$ and 
$m_{\rm max}=120\,M_\odot$, the number of stars in this subsystem is then
$N_{\rm sub} \lesssim 10^{-4}N$.
At the time of dynamical decoupling the central density of the subsystem
must be comparable to the central density of the overall cluster. Therefore we 
conclude that
the relaxation time is around three orders of magnitude smaller
for the collapsing subsystem of
heavy stars at the onset of instability. So, essentially,
the heavy stars will go into core collapse as soon as they start
dominating the mass density near the center.

If the cluster starts its evolution
with heavy stars distributed throughout, the timescale for core collapse
will be determined by their mass segregation timescale. The process of
mass segregation for the heaviest stars in the cluster is driven
by dynamical friction \citep[Section~7.1]{BT87}. The timescale for mass
segregation and the onset of core collapse therefore
depends primarily on the mass ratio between the
dominant heavy stars and the lighter background stars \citep{Spitzer69,CW90}.
For simple dynamical friction in an effective two-component model, one 
would then expect the core collapse time to be comparable to
$t_{\rm DF} \equiv (m_{\rm b}/\left< m \right>)\,t_{\rm rh}$. Here 
$m_{\rm b}$ ($\ll m_{\rm max}$, the upper mass limit of the IMF)
should be the mass above which the IMF contains a large enough number of massive stars 
to define a ``collapsing subsystem.'' If this number is $\sim 10^3$ and the
total number of stars $N\sim 10^6$ we get
$m_{\rm b}\simeq 20\,M_\odot$ for our standard Salpeter IMF and
this simple analysis would then
predict the correct order of magnitude for the core collapse time 
($t_{\rm cc}/t_{\rm rh}(0) \sim 0.1$; see Figure~\ref{fig_cc_compare}).
Note, however, that 
this simple dynamical friction picture corresponds to one or a few massive
objects traveling through a uniform background of much lighter particles, a
description that provides, at best, only a rough approximation of the real situation
under consideration here (see Appendix). 

\subsection{Goals of this Study}

In this paper, the first of a series, we consider a wide variety of initial cluster
models and we investigate the
evolution of all systems until the onset of core collapse. Results from numerical
simulations that include stellar collisions and track the growth of a massive object through
successive mergers of massive stars during core collapse will be presented in a subsequent paper.
This work is in progress and preliminary results have already been reported elsewhere
\citep{RFG03}. Here we determine
how the core collapse time is related to various initial parameters including 
the IMF (Sec.~3.2 and~4.2) and the central concentration and tidal radius of the cluster
(Sec.~3.1 and~4.3),
and we also include the possibility of initial mass segregation (Sec.~3.3 and~4.4). 
We then make a comparison with the stellar
evolution time and derive limits on cluster initial conditions to allow the
development of runaway collisions and the possible subsequent formation of a
central massive BH (Sec.~5). We also derive from our calculations an estimate of the total
mass of the stars that participate in the runaway collisions, which provides
an upper limit to the final BH mass.

We do not include stellar evolution in our calculations since our aim
here is to investigate dynamical processes taking place before even
the most massive main-sequence stars in the cluster have evolved. We
do, however, study the effects of mass loss from stellar winds and
their dependence on metallicity (Sec.~\ref{subsec_stell_evol}).  In a
subsequent paper \citep{GR03},
we will study the evolution of
``post-collapse'' star clusters in which a central IMBH is assumed to
have formed early on (through the runaway collision process). In
particular, we will study the possibility that mass loss from the
stellar evolution of the remaining massive stars (those that have
escaped the central collapse and runaway) could disrupt the cluster
\citep[cf. ][]{JNR01}, thereby producing a ``naked'' IMBH. This is
motivated by observations of ultraluminous X-ray sources in regions of
active star formation (e.g., in merging galaxies) containing many
young star clusters, but with the X-ray sources found predominantly
{\em outside\/} of those clusters \citep{ZF02}.

An important factor that could affect significantly the dynamics of core collapse 
in young star clusters is the
presence of primordial binaries \citep{FGJR03} or the dynamical formation of
binaries through three-body interactions \citep{Hut85,Giersz98}. As pointed
out by \cite{Inagaki85}, the formation rate of hard three-body binaries would
be accelerated in clusters with a mass spectrum \citep{Heggie75}.
In this first paper we do
not take into account the presence of binaries in the cluster. This is
justified because we do not expect binaries to play an important role until
after the onset of core collapse, when the central density increases
suddenly. However, collisions should become dominant immediately
after the onset of core collapse.
Our expectation, based on several previous theoretical studies of
physical collisions during interactions of hard primordial binaries and for three-body 
binary formation, is that the presence of binaries in the core will
in fact {\em increase\/} collision rates, thereby helping
to trigger the runaway 
\citep{CH96,BSD96,FGR03b}.
This is also supported by the results of direct $N$-body simulations
showing that, in smaller systems containing $N\gtrsim 10^4$
single stars, collisions indeed
occur predominantly through the interactions of 
three-body binaries formed at core collapse \citep{PZM02}. 

\section{Numerical Methods and Summary of Previous Work}
\label{sec_MC_sim}

Numerical methods for investigating the dynamical evolution of star
clusters include direct $N$-body integration, solutions of
the Fokker-Planck
equation by direct (finite-difference) or Monte Carlo methods, and
gaseous models \citep[for a review, see][]{HH03}. Here we refer to
direct $N$-body integrations simply as ``$N$-body simulations,''
direct integrations of the Fokker-Planck equation as ``Fokker-Planck
(FP) simulations,'' and Fokker-Planck simulations based on Monte Carlo techniques
as ``Monte Carlo (MC) simulations.'' Note, however, that our MC simulations, in which
the cluster is modeled on a star-by-star basis, are in fact another type of
``$N$-body simulations.'' Each approach offers different
advantages and disadvantages for understanding core collapse and
massive black hole formation in star clusters with a realistic mass
spectrum.

\subsection{Summary of Previous Numerical Work}

\cite{PZM02} carried out $N$-body simulations starting from a variety
of initial conditions for clusters containing up to $\sim 6\times
10^4$ stars. They found that runaway collisions driven by the most
massive stars can happen in sufficiently dense clusters.  Their
results apply directly to small star clusters containing $\sim
10^4-10^5$ stars.  However, in such a small cluster, any realistic IMF
typically contains only a very small number of massive stars. For
example, a Salpeter IMF with minimum mass $m_{\rm min}=0.2\,M_\odot$
and maximum mass $m_{\rm max}=120\,M_\odot$ contains a fraction
$\sim 3\times 10^{-4}$ of its stars above $60\,M_\odot$. The
dynamical role played by massive stars can therefore depend strongly
on the total number of stars in the cluster.  In addition, in small
systems, the dynamical evolution might be dominated by the random
behavior or the initial conditions of just a few very massive
stars. Consequently, to investigate runaway collisions in larger
systems such as super star clusters or proto-galactic nuclei,
realistically large numbers of stars must be used in numerical
simulations. The computational time required for direct integration of an
$N$-body system over one crossing time $t_{\rm dyn}$ scales as $N^2$
($N^{1.25}$ on parallel machines, if one can adjust the number of CPUs
to $N$ optimally; see \citealt{SB03}).  Since the relaxation time is
$\sim (N/\ln N)\,t_{\rm dyn}$ (see Sec.~\ref{CC_and_SI}), the
scaling of the total CPU time of $N$-body simulations is nearly as
steep as $N^3$. Currently, using a state-of-the-art GRAPE-6 board to
accelerate the computations \citep{Makino01,Makino02}, the evolution
of a cluster containing $10^5$ stars with $m_{\rm max}/m_{\rm
min}=1000$ and no primordial binaries can be integrated up to core
collapse in about one day (H.~Baumgardt, private
communication). However, including primordial binaries or increasing
the number of stars would still lead to prohibitively long computation
times, especially for a parameter study where a large number of
integrations are required.

A wide mass spectrum also leads to increased computation times for
FP simulations and gaseous models\footnote{We have been informed 
(H.~Cohn \& B.~Murphy 2004, private communication)
that latest isotropic FP codes require only a few seconds of computation
to reach core collapse, for as many as 20 mass components.}. In these
methods, a continuous mass spectrum is approximated by discrete mass
bins. In the gaseous model, the most time-consuming operation of the
algorithm is the inversion of large matrix whose dimension is
proportional to the number of equations, itself proportional to
$N_{\rm comp}$. Hence the computing time increases like\footnote{Note
that, in principle, one could split each step into separate
``Poisson'' and ``Fokker-Planck'' parts, in a way similar to what is
done in FP codes, hence reducing the cost to $T_{\rm CPU} \propto
N_{\rm comp}^2$ (R.~Spurzem, private communication).} $T_{\rm CPU}
\propto N_{\rm comp}^3$. For FP codes, there is one diffusion term
coupling each component to all others, leading to $T_{\rm CPU} \propto
N_{\rm comp}^2$. These steep scalings limit $N_{\rm comp}$ to at most
$\sim 20$ to avoid computation times longer than a few
days.

Even with a small number of mass bins,
FP simulations have yielded important qualitative results. Inagaki and
collaborators investigated two- and multi-component systems
\citep{IW84,Inagaki85,IS85}. They found that the energy transfer {\em
within\/} the core is an important process for determining the onset
of core collapse. FP simulations by \citet{Inagaki85} and \citet{CW90}
show that a mass function with $m_{\rm max}/m_{\rm min}\simeq 10-15$
needs to be discretized into about 15 components or more to obtain an
accurate value of the core collapse time. A coarser discretization
leads to an artificially slow evolution. \citet{CW90} successfully
tracked the energy transfer between different components and
demonstrated the importance of this process. \cite{Quinlan96} used FP
simulations to follow the evolution of single- and two-component
systems. He used a unique setting for his two-component clusters, with
each component having a different initial spatial distribution. His
aim was to study the interaction of galactic nucleus containing mainly
dark (compact) objects with the bulge of the galaxy. In one model he
considered a structure with ``inverse initial mass segregation,''
where the central nucleus is made of objects much lighter than the
normal stars composing the bulge. He showed that this situation also
leads to highly accelerated evolution as the more massive stars get
trapped by the nucleus through dynamical friction and undergo rapid
core collapse. \citet{Takahashi97} also investigated the evolution of
clusters with a mass spectrum using FP simulations. His simulations
were two-dimensional (in phase space) and therefore he was able to
study the development of velocity anisotropy.

Gaseous models have the advantage of being very fast (for $N_{\rm
comp}\la 20$) but they include the greatest number of simplifying
assumptions and so require independent checking. The good agreement
shown with our MC simulation for a Plummer model with Salpeter IMF in
Figure~\ref{fig_cc_compare} is encouraging. Note, however, that the
gaseous model calculation shown in Figure~\ref{fig_cc_compare} used
50 mass components. This number of components is exceptionally high,
leading to a total computation time in excess of three weeks, much
longer than most MC runs. It was chosen to be able to follow core
collapse to a very advanced stage, until the most massive component
dominates the collapsing core. We note that the evolution of the
gaseous model to core collapse is faster by some 30\,\% and exhibits a
more gradual contraction of the inner region. The reasons for these
small discrepancies will be investigated in future work, where
additional comparisons between Monte Carlo and gaseous model
calculations will also be presented
\citep{FASS03}. For the time being, suffice it to mention that for 
multi-mass clusters, there are basically two adjustable
(dimensionless) parameters in the gaseous model, one setting the
effective thermal conductivity and the other the timescale for energy
exchange between components ($\lambda$ and $\lambda_{\rm eq}$, see
\citealt{LS91,GS94,ST95}). For the model plotted in
Figure~\ref{fig_cc_compare}, we used the standard values of these
parameters ($\lambda=0.4977$ and $\lambda_{\rm eq}=1$) which were
established for clusters with a different structure and mass spectrum
and some adjustment (preferentially through comparisons with $N$-body
runs) may be required.

A direct comparison between FP and gaseous models was also carried out
by \cite{ST95}, for two-component clusters, and also resulted in good
agreement. Recently, a hybrid code has been developed by
\cite{GS03}, combining a gaseous model with MC techniques. In
this approach the single stars are represented by the gaseous model,
while primordial binaries are followed with a Monte Carlo treatment. A
similar hybrid treatment could be applied to the problem we are
studying here, but with massive (single and binary) stars included in
the Monte Carlo component, and lower-mass stars represented by a
gaseous model.

\subsection{The Monte Carlo Code}

The solution of the Fokker-Planck equation with an orbit-averaged Monte
Carlo method is an ideal compromise for the problem at hand. It can
handle a suitably large number of stars and a wide mass spectrum
can be implemented with very little additional difficulty. 
Most importantly, as in direct $N$-body integrations, MC simulations
can implement a star-by-star description of the cluster. This allows the 
inclusion of many important processes such as
collisions, binary interactions (including primordial binaries),
stellar evolution (and the accompanying mass loss), as well as the effects of a
massive central object with relative ease 
and much higher realism  compared to direct FP simulations
or gaseous models. We will incorporate the effects of all these additional
processes on cluster evolution in the subsequent papers of this series.

The MC code we have used to obtain the main results
of this paper is described in detail by \cite{JRPZ00}. It is based on the
ideas of \cite{Henon71,Henon71b,Henon73} and in many respects it is very
similar to MC codes developed by \cite{Stodol82,Stodol86},
\cite{Giersz98,Giersz01},
and \cite{FB01}. In the rest of this section, we give a brief summary
of the numerical method.

The main simplifying assumption is the Fokker-Planck approximation, in which 
relaxation processes
are assumed to be dominated by small distant encounters rather than strong
encounters with large deflections \citep{Spitzer87,BT87}. The 
dynamical evolution can then be
treated as a diffusion process in phase space. Following the individual
interactions between the stars, as in direct $N$-body simulations, is
computationally expensive. However, the {\em average\/} cumulative effect on
a star in a given amount of time can be characterized by
diffusion coefficients in phase space.
To compute the relaxation, the timestep for the numerical evolution has to be
chosen smaller than the relaxation time.
Since the relaxation time is normally shortest at the center, we choose
our timestep to be a fraction of the central relaxation time. This ensures that
the relaxation is followed accurately throughout the cluster.

Another important simplification in the MC method is the assumption of
spherical symmetry. The position of the particles is represented by a single
radial coordinate, $r$, and the velocity is represented by radial, $v_r$,
and tangential, $v_t$, components. The specific angular momentum, $A$,
and the specific energy, $E$, of a star with index $i$, are given by:
\begin{equation}
A_i=v_{t,i}\, {r_i}\:,\quad E_i=U(r_i)+\frac{1}{2}
(v_{r,i}^2+v_{t,i}^2),
\end{equation}
where $U(r)$ is the gravitational potential at a given point. The assumption
of spherical symmetry implies that the potential at all points
can be computed in a time proportional
to the number of particles $N$, rather than $N^2$.

In H\'enon's algorithm, the evolution is simulated by
reproducing the effect of the cluster on each star by a single effective
encounter in each timestep. 
At every iteration the two integrals of the motion
$A_i$ and $E_i$ characterizing the orbit of each
star are perturbed in a way that is consistent with the value of the
diffusion
coefficients \citep{Henon73}. 
To conserve energy, this perturbation is realized by a single
effective scattering
between two neighboring stars. The square of the scattering angle, $\beta$,
is proportional to the timestep chosen. We choose our timestep such
that $\sin^2(\beta/2) \lesssim 0.05$ at the center of the cluster.
Choosing too large a timestep will lead to a saturation effect and the
relaxation will proceed artificially slowly. Using too small a timestep, on
the other
hand, not only increases the computation time, but also can lead to spurious
relaxation (see the Appendix for a discussion of spurious relaxation effects). 

After the perturbation of $A_i$ and $E_i$,
the stars are placed at random positions between
their apocenters and pericenters using a probability distribution that is
proportional to the time spent at a given location on their new orbit.
For a star with index $i$, the apocenter and pericenter distances are
calculated by finding the roots of
\begin{equation}
2E_i-2U(r)-\frac{A_i^2}{r^2} = 0.
\end{equation}
This random placement is justified by the assumption of dynamical equilibrium,
i.e., the evolution of the system does not take place on the crossing
(or dynamical) timescale, but rather on the relaxation timescale. The only important 
point about assigning a specific position
to a particle on its orbit is that its contribution to density, potential, interaction
rates, etc., has to be estimated correctly. 
After all stars have been placed at their new positions,
the potential is recalculated and the whole cycle of perturbation is repeated.

This method can be modified so that the timestep is a fraction of
the {\em local\/} relaxation time \citep{Henon73,FB01}. \cite{Stodol82,Stodol86}
and \cite{Giersz98,Giersz01} divided the system into zones resulting in
an approach intermediate between a fixed and smoothly varying timestep.
Dividing the system into radial zones also allows the MC method to be parallelized
efficiently for use on multi-processor machines \citep{JRPZ00}.
Another possible modification uses a scaling of the units such
that each particle in the simulation can represent an entire spherical
shell of many identical stars rather than a single star 
\citep{Henon71,FB01,FB02}. 
It should also be noted
that, although the effects of strong encounters between stars on relaxation
are assumed to be negligible compared to weak, more distant encounters,
they can be incorporated by estimating their rate of occurrence in a way
similar to physical collisions \citep{FB02} or interactions with binary
stars \citep{FGJR03}.

Our MC code has been used previously to study 
many fundamental dynamical processes
such as the Spitzer instability \citep{WJR00} and mass segregation
\citep{FJPZR02} for simple two-component systems, as well as the evolution
of systems with a continuous but fairly narrow mass spectrum of evolving stars
\citep{JNR01}.  A difficulty introduced by a broad continuous IMF (with a
large $m_{\rm max}/m_{\rm min}$ ratio) is the necessity of adjusting
the timestep to treat correctly encounters between stars of very
different masses. When pairs of stars are selected to undergo an
effective hyperbolic encounter as described above, one has to make
sure that the deflection angle remains small for {\em both\/}
stars. In situations where the mass ratio of the pair can be extreme,
one has to decrease the timestep accordingly \citep{Stodol82}.  In practice, for the
simulations described here, we find that the timestep has to be
reduced by a factor of up to $\sim500$ compared to what would be
appropriate for a cluster of equal-mass stars. We discuss further the
applicability of orbit-averaged MC methods to systems with a
continuous mass spectrum and large $m_{\rm max}/m_{\rm min}$ ratio in
the Appendix.

\section{Initial Conditions and Units}
\label{init_cond}

The characteristics of core collapse and the subsequent runaway collisions
depend on the initial conditions for the cluster. These initial conditions
include the total number of stars, the IMF, the initial spatial distribution
of the 
stars (density profile and, possibly, initial mass segregation), and the
position of the
cluster in the galaxy, which determines the tidal boundary. As we shall see,
the most important 
initial parameters are the slope of the IMF, the maximum stellar mass, and
the initial
degree of central concentration of the cluster density profile.
We have used a wide variety of initial conditions, both to test the
robustness
of our findings and to establish the dependence of our results on these
parameters. As a typical reference model we use an isolated Plummer sphere
with a 
Salpeter IMF and stellar masses ranging from $m_{\rm min}=0.2\,M_\odot$ to
$m_{\rm max}=120\,M_\odot$. We then explore variations on this model by
changing
the initial cluster structure, the IMF or the number of stars.

\subsection{Density Profile}

\begin{figure}
\resizebox{\hsize}{!}{\includegraphics[clip]{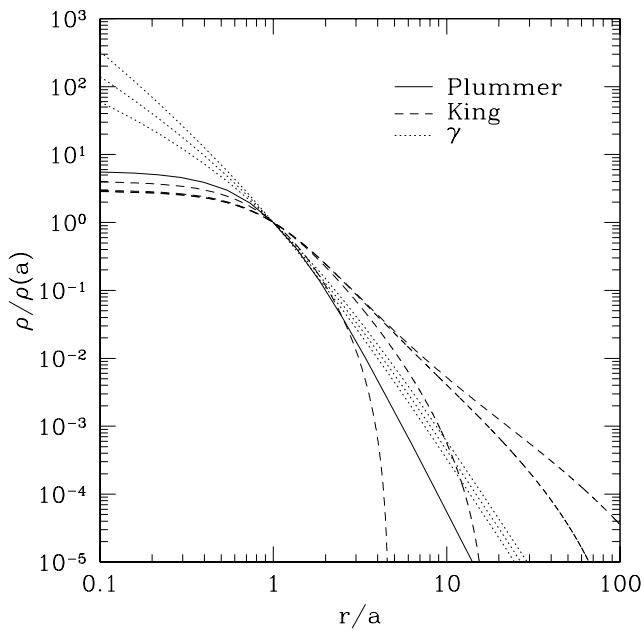}}
\caption{
Density profiles for various initial models used in our simulations. 
The Plummer model is shown by a solid line,
$\gamma$-models corresponding to $\gamma=1, \,1.5$ and~2 are shown by
dotted lines, and King models with $W_0=3,\,6,\,9$ and~12 are
shown by dashed lines. The radius is in units of the characteristic
length scale parameter of each model and the density profiles are
normalized to unity at that radius. King models with lower $W_0$
values have steeper profiles outside $r=a_{\rm K}$, while
$\gamma$-models with higher $\gamma$ values have higher $\rho/\rho(a)$
near the center. 
\label{fig_dens_prof}
}
\end{figure}

We have examined three families of models:
Plummer and King models \citep[Sections 2.2 and 4.4]{BT87}, which have a
core-halo structure,
and $\gamma$-models \citep{Dehnen93,Tetal94}, which have a cusp near the center.
All these models
have a characteristic radius given by
\begin{align}
a_{\rm P} &= (2^{2/3}-1)^{1/2} r_{\rm h} \simeq 0.766 \, r_{\rm h}, \label{eq_aPlummer}\\
a_{\rm K} &= \sqrt{ \frac{9\sigma^2}{4\pi\,G\,\rho_0}},\\
a_\gamma &= (2^{1/(3-\gamma)}-1)r_{\rm h},\label{eq_aGamma}
\end{align}
for Plummer, King, and $\gamma$-models respectively. In these
formulae, $r_{\rm h}$ is the half-mass radius, $\rho_0$ is the central
density, and $\sigma$ is a King model parameter.  We show the density
profiles corresponding to these various models in
Figure~\ref{fig_dens_prof}. Here $W_0$ is the dimensionless central
potential, related to the concentration parameter 
\citep[Fig.~4-10]{BT87}. 
Other useful
quantities characterizing the various density profiles are given in
Table~\ref{table_clust}. The initial model used for each of our
MC simulations is listed in the second column of Table~\ref{table_models}.
A general procedure for producing these models is given by
\cite{FB02}; a less general but simpler procedure for the Plummer
model is given by \cite{AHW74}. 
\begin{deluxetable*}{rrllllllll}
\tabletypesize{\scriptsize}
\tablecaption{Properties of clusters
\label{table_clust}}
\tablewidth{0pt}
\tablehead{
\colhead{Cluster} & \colhead{$\rho_{0}$} & \colhead{$\sigma_{0}$} & \colhead{$a$} & \colhead{$r_{\rm t}$} & \colhead{$r_{\rm h}$}  & \colhead{$r_{\rm c}$}  & \colhead{$M_{\rm c}$} & \colhead{$t_{\rm rh}$} & \colhead{$t_{\rm rc}$}
}
\startdata
Plummer       &  1.167 & 0.532   & 0.589  & $\infty$ & 0.769 & 0.417  & 0.192  & 0.093  & 0.0437  \\
\tableline                                                            			         
 King $W_0=1$ &  0.454 & 0.534   & 1.517  & 2.568    & 0.858 & 0.670  & 0.321  & 0.110  & 0.1134  \\
 2            &  0.530 & 0.526   & 1.003  & 2.800    & 0.849 & 0.612  & 0.281  & 0.108  & 0.0930  \\
 3            &  0.652 & 0.518   & 0.749  & 3.134    & 0.839 & 0.543  & 0.238  & 0.106  & 0.0722  \\
 4            &  0.860 & 0.510   & 0.576  & 3.625    & 0.827 & 0.465  & 0.195  & 0.104  & 0.0523  \\
 5            &  1.252 & 0.504   & 0.438  & 4.362    & 0.814 & 0.382  & 0.1546 & 0.101  & 0.0348  \\
 6            &  2.112 & 0.503   & 0.320  & 5.471    & 0.804 & 0.293  & 0.1171 & 0.100  & 0.0205  \\
 7            &  4.526 & 0.511   & 0.2146 & 6.987    & 0.812 & 0.2032 & 0.0830 & 0.101  & 0.00997 \\
 8            & 13.742 & 0.530   & 0.1253 & 8.344    & 0.872 & 0.1211 & 0.0531 & 0.112  & 0.00368 \\
 9            & 55.671 & 0.558   & 0.0649 & 8.374    & 0.980 & 0.0633 & 0.0307 & 0.134  & 0.00106 \\
\tableline                                                            
$\gamma=1.0$  &$\infty$& 0       & $1/3$  & $\infty$ & 0.805 & 0      & 0      & 0.100  & 0        \\
       $1.5$  &$\infty$& 0       & $1/2$  & $\infty$ & 0.851 & 0      & 0      & 0.108  & 0        
\enddata
\tablecomments{Definitions of quantities listed in this table:
$\rho_{0}$ is the central mass density; $\sigma_{0}$ is the central
one-dimensional velocity dispersion; $a$ is the characteristic radius
(see Eqs.~\ref{eq_aPlummer}--\ref{eq_aGamma}); $r_{\rm t}$ is the
tidal radius; $r_{\rm h}$ is the half-mass radius; $r_{\rm c}\simeq
\left[9\sigma_0^2/(4\pi G\rho_0)\right]^{1/2}$ is the core radius;
$M_{\rm c}$ is the mass enclosed by $r_{\rm c}$; $t_{\rm rh}$ is the
half-mass relaxation time (eq.~\ref{eq_rel_time}); and $t_{\rm rc}$ is
the central relaxation time (eq.~\ref{eq_trc}). All quantities are
given in $N$-body units, except that $t_{\rm rh}$ and $t_{\rm rc}$ are
given in ``Fokker-Planck units'' (see Section~\ref{sec_units}).
}
\label{table_cluster_prop}
\end{deluxetable*}

\subsection{Initial Mass Function}
\label{subsec_mass_fcts}

Since we expect the conditions leading to core collapse to depend
sensitively  
on the IMF, we have carried out simulations with a wide range of IMF
parameters. However, it is generally established that, at least for the
high-mass end of the spectrum, a universal IMF very close to the simple
Salpeter-like power-law is obeyed. This is indicated
both by observations and by theoretical calculations
\cite[and references therein]{Kroupa02,Clarke03}.

In our simulations we assign the masses
of individual stars using a sampling procedure.
We first choose a random number, $X$,
from a uniform distribution between 0 and 1. For a simple power-law IMF with
$dN \propto  m^{-\alpha}\,dm$ between $m_{\rm min}$ and $m_{\rm max}$ we
calculate the corresponding mass using
\begin{equation}
\label{eq_Salpeter}
m(X) = m_{\rm min}\left(1+X
 \left[\left(\frac{m_{\rm max}}{m_{\rm min}}\right)^{1-\alpha}-1\right]
\right)^{1/(1-\alpha)},
\end{equation}
where the value $\alpha = 2.35$ would correspond to a Salpeter IMF.

In addition to simple power laws, for two models we have used the
Miller-Scalo \citeyearpar{MS79} and Kroupa \citep{KTG93} IMFs, which
are steeper at the high-mass end of the spectrum and shallower at the
low-mass end. Both of these IMFs can be represented by broken power laws. 
For the Miller-Scalo IMF,
\begin{equation}
\frac{dN}{dm} \propto \begin{cases}
  m^{-1.4} \; & 0.1  \leq m \leq 1.0 \\
  m^{-2.5} \; & 1.0  \leq m \leq 10.0 \\
  m^{-3.3} \; & 10.0 \leq m
\end{cases} ,
\end{equation}
and for the Kroupa IMF,
\begin{equation}
\frac{dN}{dm} \propto \begin{cases}
  m^{-1.3} \; & 0.08 \leq m \leq 0.5 \\
  m^{-2.2} \; & 0.5  \leq m \leq 1.0 \\
  m^{-2.7} \; & 1.0 \leq m
\end{cases} ,
\end{equation}
where all numerical values are in solar mass.

However, for the sake of computational convenience, we prefer
to implement a somewhat different parametrization of these
distributions. To generate the mass spectra corresponding to these
IMFs we use the functions \citep{ETF89,KTG93}
\begin{equation}
\label{eq_MillerScalo}
m(X) = \frac{0.19 X}{(1-X)^{0.75} + 0.032 (1-X)^{0.25}}
\end{equation}
and
\begin{equation}
\label{eq_Kroupa}
m(X) = 0.08 + \frac{0.19X^{1.55}+0.05 X^{0.6}}{(1-X)^{0.58}},
\end{equation}
for the Miller-Scalo and Kroupa mass functions, respectively. In both
cases if $m(X)<m_{\rm min}$ or $m(X)>m_{\rm max}$ the result is
discarded and a new random number is generated. The values in
equation~(\ref{eq_Kroupa}) correspond to the choice of $\alpha_1=1.3$
in Table~10 of \cite{KTG93}. 

In most of our simulations we have used $N = 1.25\times10^6$
stars. For all mass functions, this implies the presence of many
massive stars, allowing us to resolve fully the higher end of the mass
spectrum.  For example, for $m_{\rm min}=0.2\,M_\odot$ and $m_{\rm
max}=120\,M_\odot$, with $N=1.25\times10^6$, we have $\gtrsim 60$ stars
with $m>100\,M_\odot$ for a Salpeter IMF.  The results from our
simulations are therefore not much affected by random fluctuations in
a small number of very massive stars. Note, however, that in much
smaller systems, containing perhaps only $\sim 10^4$ stars (as in the
Arches and Quintuplet clusters near our Galactic center), these small
number effects and random fluctuations may indeed play a dominant role in determining the
dynamical fate of the few most massive stars in the system.

\subsection{Initial Mass Segregation}

Initial mass segregation in star clusters (i.e., the tendency for
more massive stars to be {\em formed\/} preferentially near the
cluster center) is expected to result from star formation feedback in
dense gas clouds \citep{ML96} or from competitive gas accretion onto
proto-stars and mergers between them \citep{BBCP01,BB02}.  There is also
some observational evidence for initial mass segregation in both open and
globular clusters \citep{BD98,RM98,dGGJ03}. 

We have considered the possibility of
initial mass segregation in a few of our MC simulations. 
We adopt a simple prescription
whereby we increase the average stellar
mass within a certain radius $r_{\rm ms}$.
For $r<r_{\rm ms}$, rather than sampling from an IMF with fixed $m_{\rm
min}$,
we randomly choose between two values of $m_{\rm min}$, one that is used for
the outer part of the cluster and another one that is larger. For
$r>r_{\rm ms}$, we follow a similar procedure, this time changing  
$m_{\rm max}$. The average mass within $r_{\rm ms}$ is larger by a factor
$C_{\rm ms}$ and outside $r_{\rm ms}$ is smaller by a factor 
$C_{\rm ms}'$, with respect to a cluster without initial mass segregation.
We choose $C_{\rm ms}$ and $C_{\rm ms}'$ such that the overall average stellar mass
in the cluster does not change with these modifications.
Changing the average stellar mass within any region of the cluster
would of course in general leave the system out of dynamical equilibrium.
To maintain virial equilibrium, the mass density profile must also be preserved. 
We achieve this by modifying the number density of stars appropriately. 

Our initial conditions for models with initial mass segregation
are summarized in Table~\ref{table_ims}. Here $q$ is the initial cluster mass 
fraction within $r_{\rm ms}$. This implies $C_{\rm ms}' = (1-q)/(1 - q C_{\rm ms})$. 
Increasing $q$ or $C_{\rm ms}$
represents more extended or more pronounced initial mass segregation.

Our prescription for initial mass segregation allows the formation
of massive stars in the outer parts of the cluster,
as well as the formation of lighter stars in the inner
parts, but the more massive stars are more likely to be found near the
center.  This makes our approach different from that of \citet{BD98},
who put all massive stars closer to center. 

\subsection{Units}
\label{sec_units}

For all our numerical calculations, we adopt the standard $N$-body units
\citep{Henon71}: we set
the initial total cluster mass $M=1$, the gravitational constant $G=1$, and the
initial total energy $E_0=-1/4$. In the tabulation of the results we
also use the initial half-mass relaxation time $t_{\rm rh}(0)$,
given by equation~(\ref{eq_rel_time}), as the unit of time for
comparison with other work in the literature.

The conversion to physical units is done
by evaluating the initial half-mass relaxation
time in years. For example, for the Plummer model, we can write
\begin{multline}
\label{eq_conv}
t_{\rm rh}(0) \simeq 330\,{\rm Myr} \\
\times
\left(\frac{N/10^6}{\ln (\gamma_c N) / \ln (10^4)}\right)
\left(\frac{a_{\rm P}}{1\,{\rm pc}}\right)^{3/2}
\left(\frac{M}{10^6 M_\odot}\right)^{-1/2} .
\end{multline}
Similar expressions can also be obtained for King models and
$\gamma$-models by use of the quantities in
Table~\ref{table_cluster_prop}.
For a single-component model (containing equal-mass stars) the value of $\gamma_{\rm
c}$ in the Coulomb logarithm can be calculated theoretically \citep{FS94},
but for a
system with a wide mass spectrum it must be determined by
comparing to direct $N$-body integrations. \cite{GH96,GH97} carried
out such comparisons and found $\gamma_{\rm c}\simeq0.015$.  Our own
comparison with a recent $N$-body result for $\gtrsim 10^5$ stars with
a wide mass spectrum led us to adopt the value $\gamma_{\rm c} = 0.01$
(H.~Baumgardt, private communication).

For processes occurring in the central parts  of the cluster, the central
relaxation time is a more relevant quantity,
\begin{equation}
\label{eq_trc}
t_\mathrm{rc}(0)\equiv \frac{ \sigma_{\rm 3D}^3 }{ 4.88\pi \,G^2 \ln(\gamma_\mathrm{c}N)\,n
\langle m \rangle^2 } ,
\end{equation}
where $\sigma_{\rm 3D}$, $n$ and $\langle m
\rangle$ are the 3D velocity dispersion, number density and average
stellar mass at the cluster center \citep[][eq.~3.37]{Spitzer87}.
The $N$-body unit system only specifies unambiguously 
{\em dynamical times\/}, which are independent of $N$. In this system,
relaxation times are proportional to $N/\ln(\gamma_\mathrm{c}N)$. It
is therefore useful to define also the so-called ``Fokker-Planck time
unit,'' which is the $N$-body time unit ($t_{\rm dyn}\equiv GM^{5/2}(-4E_0)^{-3/2}$)
multiplied by $N/\ln(\gamma_\mathrm{c}N)$.

The most important physical properties of all our initial cluster
models, including $t_{\rm rh}(0)$ and $t_{\rm rc}(0)$, are given in
Table~\ref{table_cluster_prop} (in $N$-body and FP units).

\section{Results}
\begin{deluxetable*}{rllrrrr} 
\tabletypesize{\scriptsize}
\tablecaption{Initial conditions and main results
\label{table_models}}
\tablewidth{0pt}
\tablehead{
\colhead{Model} & \colhead{Initial Structure} & \colhead{IMF} & 
\colhead{$\tbfrac{m_{\rm min}\text{--}m_{\rm max}}{(M_\sun)}$} & 
\colhead{$\tbfrac{\left<m\right>}{(M_\sun)}$} & 
\colhead{$t_{\rm cc}/t_{\rm rh}(0)$} & 
\colhead{$M_{\rm cc}/M_{\rm tot}$} 
}
\startdata
 55 & Plummer        & PL, $\alpha=-1.4$  & 0.2--120  & 6.57 & 0.287   & $\cdots$ \\
 50 & Plummer        & PL, $\alpha=-1.7$  & 0.2--120  & 2.74 & 0.131   & 0.0024 \\
  1 & Plummer        & PL, $\alpha=-2.0$  & 0.2--120  & 1.28 & 0.0899  & 0.0018 \\
 51 & Plummer        & PL, $\alpha=-2.2$  & 0.2--120  & 0.87 & 0.0700  & 0.0020 \\
 2r & Plummer        & PL, $\alpha=-2.35$ & 0.2--120  & 0.69 & 0.0716$^{\rm a}$  & $\cdots$ \\
 2s & Plummer        & PL, $\alpha=-2.35$ & 0.2--120  & 0.69 & 0.0702$^{\rm a}$  & $\cdots$ \\
  2 & Plummer        & PL, $\alpha=-2.35$ & 0.2--120  & 0.69 & 0.0700$^{\rm b}$  & 0.0020$^{\rm b}$ \\
 2b & Plummer        & PL, $\alpha=-2.35$ & 0.2--120  & 0.69 & 0.0700$^{\rm b}$  & 0.0019$^{\rm b}$ \\
 2c & Plummer        & PL, $\alpha=-2.35$ & 0.2--120  & 0.69 & 0.0706  & 0.0020 \\
 2d & Plummer        & PL, $\alpha=-2.35$ & 0.2--120  & 0.69 & 0.0720  & 0.0018 \\
 52 & Plummer        & PL, $\alpha=-2.5$  & 0.2--120  & 0.58 & 0.0719  & 0.0022 \\
  3 & Plummer        & PL, $\alpha=-2.7$  & 0.2--120  & 0.48 & 0.0696  & 0.0018 \\
 53 & Plummer        & PL, $\alpha=-3.0$  & 0.2--120  & 0.40 & 0.0834  & 0.0012 \\
  4 & Plummer        & Kroupa             & 0.2--120  & 0.96 & 0.0858  & 0.0014 \\
  5 & Plummer        & Miller-Scalo       & 0.2--120  & 0.71 & 0.0723  & 0.0022 \\
 28 & Plummer        & PL, $\alpha=-2.35$ & 0.2--360  & 0.72 & 0.0795  & 0.0020 \\
 27 & Plummer        & PL, $\alpha=-2.35$ & 0.2--240  & 0.71 & 0.0760  & 0.0020 \\
 24 & Plummer        & PL, $\alpha=-2.35$ & 0.2--90   & 0.68 & 0.0664  & 0.0014 \\
 20 & Plummer        & PL, $\alpha=-2.35$ & 0.2--60   & 0.67 & 0.0786  & 0.0024 \\
 21 & Plummer        & PL, $\alpha=-2.35$ & 0.2--20   & 0.62 & 0.156   & 0.0028 \\
 22 & Plummer        & PL, $\alpha=-2.35$ & 0.2--8    & 0.56 & 0.478   & 0.0010 \\
 25 & Plummer        & PL, $\alpha=-2.35$ & 0.2--5    & 0.53 & 0.805   & 0.0016 \\
 23 & Plummer        & PL, $\alpha=-2.35$ & 0.2--2    & 0.45 & 2.20    & $\cdots$ \\
 26 & Plummer        & PL, $\alpha=-2.35$ & 0.2--1    & 0.37 & 4.29    & $\cdots$ \\
 30 & King, $W_0=1$, (i) & PL, $\alpha=-2.35$ & 0.2--120  & 0.69 & 0.152   & 0.0020 \\
 36 & King, $W_0=1$      & PL, $\alpha=-2.35$ & 0.2--120  & 0.69 & 0.151   & 0.0014 \\
 35 & King, $W_0=2$, (i) & PL, $\alpha=-2.35$ & 0.2--120  & 0.69 & 0.134   & 0.0020 \\
 37 & King, $W_0=2$      & PL, $\alpha=-2.35$ & 0.2--120  & 0.69 & 0.129   & 0.0014 \\
 11 & King, $W_0=3$, (i) & PL, $\alpha=-2.35$ & 0.2--120  & 0.69 & 0.107   & 0.0022 \\
 10 & King, $W_0=3$      & PL, $\alpha=-2.35$ & 0.2--120  & 0.69 & 0.110   & 0.0016 \\
 31 & King, $W_0=4$, (i) & PL, $\alpha=-2.35$ & 0.2--120  & 0.69 & 0.0779  & 0.0022 \\
 38 & King, $W_0=4$      & PL, $\alpha=-2.35$ & 0.2--120  & 0.69 & 0.0778  & 0.0022 \\
 32 & King, $W_0=5$, (i) & PL, $\alpha=-2.35$ & 0.2--120  & 0.69 & 0.0561  & 0.0024 \\
 39 & King, $W_0=5$      & PL, $\alpha=-2.35$ & 0.2--120  & 0.69 & 0.0526  & 0.0020 \\
 12 & King, $W_0=6$, (i) & PL, $\alpha=-2.35$ & 0.2--120  & 0.69 & 0.0336  & 0.0014 \\
 40 & King, $W_0=6$      & PL, $\alpha=-2.35$ & 0.2--120  & 0.69 & 0.0322  & 0.0020 \\
 33 & King, $W_0=7$, (i) & PL, $\alpha=-2.35$ & 0.2--120  & 0.69 & 0.0163  & 0.0014 \\
 41 & King, $W_0=7$      & PL, $\alpha=-2.35$ & 0.2--120  & 0.69 & 0.0150  & 0.0018 \\
 34 & King, $W_0=8$, (i) & PL, $\alpha=-2.35$ & 0.2--120  & 0.69 & 0.00545 & 0.0010 \\
 42 & King, $W_0=8$      & PL, $\alpha=-2.35$ & 0.2--120  & 0.69 & 0.00577 & 0.0012 \\
 13 & King, $W_0=9$, (i) & PL, $\alpha=-2.35$ & 0.2--120  & 0.69 & 0.00135 & 0.0010 \\
 43 & King, $W_0=9$      & PL, $\alpha=-2.35$ & 0.2--120  & 0.69 & 0.00138 & 0.0012 \\
  6 & $\gamma=1$, Hernquist & PL, $\alpha=-2.35$ & 0.2--120   & 0.69 & $<10^{-4}$  & $\cdots$ \\
  7 & $\gamma=1.5$       & PL, $\alpha=-2.35$ & 0.2--120   & 0.69 & $<2\times10^{-6}$  & $\cdots$ \\
\enddata
\tablenotetext{a}{These results are obtained by averaging over 20 runs.}
\tablenotetext{b}{These results are obtained by averaging over 10 runs.}
\tablecomments{All models 
have $N=1.25\times 10^6$ stars except for Model~2r ($N=3\times10^5$), Model~2s ($N=6\times10^5$),
Model~2b ($N=2.5\times10^6$), Model~2c ($N=5\times 10^6$), and Model~2d ($N=10^7$). 
Isolated King models are indicated by (i). When an entry is missing in the last column ($\cdots$),
we were not able to determine $M_{\rm cc}$ reliably for that model.
}
\end{deluxetable*}
\begin{deluxetable}{rlrrrrrrr}
\tabletypesize{\scriptsize}
\tablecaption{Initial conditions and results of simulations with
initial mass segregation.
\label{table_ims}}
\tablewidth{0pt}
\tablehead{
\colhead{Model} & $q$ & $r_{\rm ms}/r_{\rm rh}(0)$ & $C_{\rm ms}$ & $C'_{\rm ms}$ &
\colhead{$t_{\rm cc}/t_{\rm rh}(0)$}     & \colhead{$M_{\rm cc}/M_{\rm tot}$}
}
\startdata
 m01 & 0.3 & 0.69 & 1.2 & 1.094 & 0.0588 & 0.0018 \\
 m17 & 0.3 & 0.69 & 1.5 & 1.273 & 0.0490 & 0.0020 \\
 m02 & 0.3 & 0.69 & 1.8 & 1.522 & 0.0443 & 0.0022 \\
 m18 & 0.3 & 0.69 & 2.1 & 1.892 & 0.0366 & 0.0022 \\
 m04 & 0.2 & 0.55 & 1.2 & 1.053 & 0.0637 & 0.0030 \\
 m15 & 0.2 & 0.55 & 1.5 & 1.143 & 0.0512 & 0.0026 \\
 m05 & 0.2 & 0.55 & 1.8 & 1.250 & 0.0498 & 0.0020 \\
 m16 & 0.2 & 0.55 & 2.1 & 1.379 & 0.0439 & 0.0022 \\
 m06 & 0.2 & 0.55 & 2.4 & 1.538 & 0.0399 & 0.0022 \\
 m08 & 0.1 & 0.40 & 1.2 & 1.023 & 0.0664 & 0.0016 \\
 m13 & 0.1 & 0.40 & 1.5 & 1.059 & 0.0588 & 0.0018 \\
 m09 & 0.1 & 0.40 & 1.8 & 1.098 & 0.0558 & 0.0016 \\
 m14 & 0.1 & 0.40 & 2.1 & 1.139 & 0.0560 & 0.0020 \\
 m10 & 0.1 & 0.40 & 2.4 & 1.184 & 0.0519 & 0.0018 \\
 m11 & 0.1 & 0.40 & 3.0 & 1.286 & 0.0506 & 0.0022 \\
 m12 & 0.1 & 0.40 & 3.6 & 1.406 & 0.0471 & 0.0016 \\
\enddata
\tablecomments{All initial models are isolated Plummer spheres containing
$1.25\times 10^6$ stars. Here $q$ is the mass fraction initially contained
within $r_{\rm ms}$. Inside this radius,
the average stellar mass is larger by a factor $C_{\rm ms}$ 
compared to Model~2. Outside this radius the average stellar mass is smaller
 by a factor $C'_{\rm ms}$. See Sec.~3.3 for details. 
}
\end{deluxetable}

\label{sec_results}

The initial conditions and main results of all our MC simulations are
summarized in 
Tables~\ref{table_models} and~\ref{table_ims}. All models have initially
$N = 1.25\times10^6$ stars, except for Models~2r, s, b, c, and d, which have
varying $N$ between $3\times 10^5$ and $10^7$. The maximum value of
$N$ is set in practice by the available computer memory and $N\sim 10^7$ 
corresponds to 2 GB of available memory for our code. The run for Model~2d took about
two weeks of CPU time to complete on a 2.8~GHz Pentium~4 Linux workstation.
More typical runs for
$N=1.25\times10^6$ took around $20-30$ CPU hours. The close
agreement between the outcomes of Models~2r -- 2d confirms the expectation that
our results should be independent of the
number of stars in the system, at least for sufficiently large $N$ to avoid
small-number effects in the cluster core.

\subsection{Mass Segregation and Core Collapse}
\label{sec_ms_cc}

\begin{figure}
\resizebox{\hsize}{!}{\includegraphics[clip]{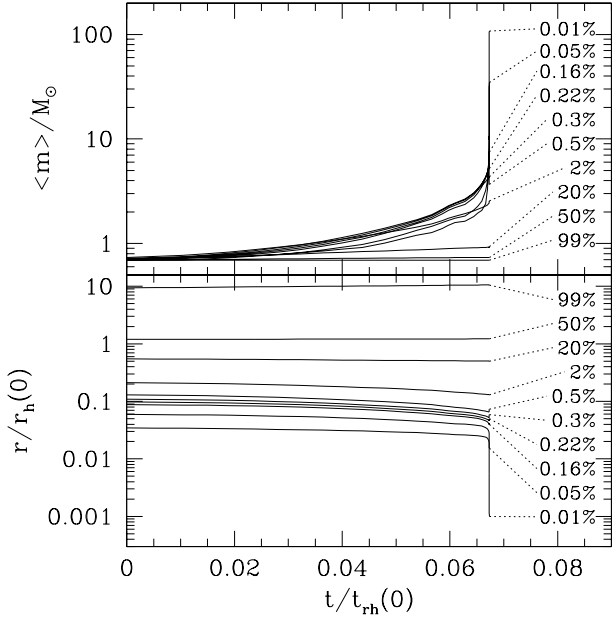}}
\caption{
Evolution of the Lagrange radii (bottom) and the average mass within
Lagrange radii (top)
for Model~2. The Lagrange radii are given in units of the initial half-mass
radius and time in units of the initial half-mass relaxation
time. A more detailed view concentrating on core collapse is given
in Figure~\ref{fig_model2_zoom}.
\label{fig_model2}
}
\end{figure}

\begin{figure}
\resizebox{\hsize}{!}{\includegraphics[clip]{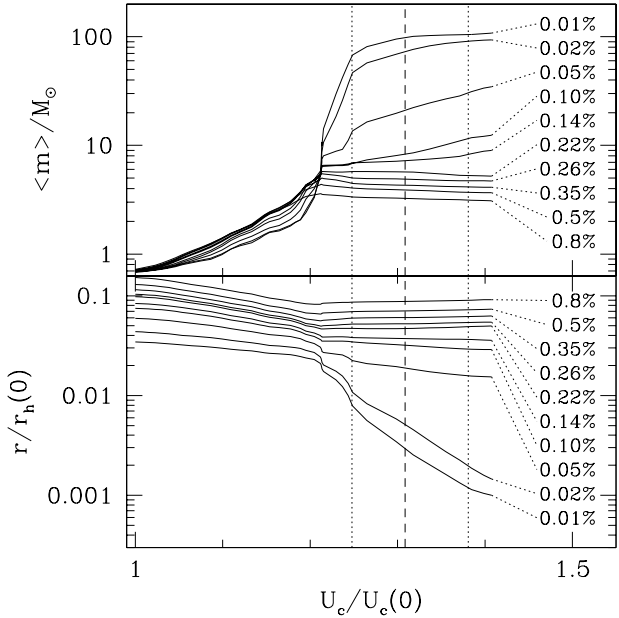}}
\caption{
Same as Figure~\ref{fig_model2} but concentrating on the evolution of
the cluster near core collapse. The horizontal axis now gives the
value of the central potential, normalized to its initial value. The
vertical dashed line marks the core collapse time, $t_{\rm cc}/t_{\rm
rh}(0) = 0.068$.  Dotted lines on either side indicate a $\pm 0.01$\%
change in this quantity, illustrating how precisely the onset of core
collapse can be determined for a particular run. The corresponding
mass fraction in the collapsing core is $M_{\rm cc}/M_{\rm tot} =
0.0018$ for this run.
\label{fig_model2_zoom}
}
\end{figure}

As expected, all models with a Salpeter-like IMF and a wide mass spectrum
undergo core collapse considerably
faster than any single-component cluster (cf.\ Sec.~\ref{CC_and_SI}).
In Figure~\ref{fig_model2}, we
show the evolution of various Lagrange
radii, as well as the average stellar mass inside these radii,
for our reference model (Model~2; same as shown in
Figure~\ref{fig_cc_compare}).
In contrast 
to the evolution of a single-component model, the inner Lagrange radii
remain almost constant until the very abrupt onset of core collapse
at $t/t_{\rm rh}(0)\simeq 0.07$.
A more detailed view of core collapse is shown in
Figure~\ref{fig_model2_zoom}. Here the time axis has been replaced by the
central potential depth, which increases monotonically in time and provides
a natural stretch near core collapse. Note that the core collapse time for a
particular run can be determined very accurately, to within $\sim 0.01\%$.
However, random fluctuations in the realization of each initial condition for a
particular model lead to a much larger physical ``error bar'' on $t_{\rm cc}$ (see below).

\begin{figure}
\resizebox{\hsize}{!}{\includegraphics[clip]{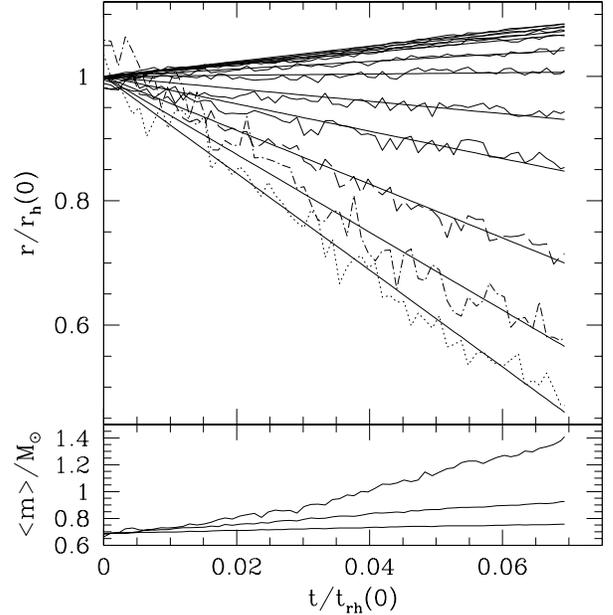}}
\caption{
Mass segregation in Model~2.
In the upper panel we show the evolution of the half-mass radii of stars in
various mass bins.
From top to bottom, the limits of the mass bins, in solar mass, are 0.2, 0.32,
0.55, 0.94, 1.62, 2.77, 4.7, 8.1, 14, 30, 50, and 120.
In the lower panel we show the average stellar mass within, from top to
bottom, $r_{\rm h}/4$, $r_{\rm h}/2$, and $r_{\rm h}$.
\label{fig_mass_segre}
}
\end{figure}

The rapid increase of the average stellar mass inside the innermost
Lagrange radii seen in Figures~\ref{fig_model2}
and~\ref{fig_model2_zoom} is an indication of significant mass
segregation. In fact, very significant mass segregation takes place
throughout the evolution of the system. This can be seen in the upper
panel of Figure~\ref{fig_mass_segre}, which shows the evolution of
half-mass radii for stars in various mass bins \citep[compare with
Fig.~1 of][]{SS75}.  It is clear that the rate of mass segregation
in each mass bin, measured by the slope $\alpha_{\rm ms}$ of the
corresponding half-mass radius $r(t)$ in that bin, is very nearly
constant from $t=0$ all the way to core collapse (the
least-square
straight-line fits, constrained to $r/r_{\rm h}(0)=1$ at $t=0$,
are
shown in the figure). This rate can be positive or negative. For
Model~2 we find that stars more
massive than about $5\,M_\odot$ drift
inward on average, while
less massive stars drift outward. Even for
the most massive stars, we do {\em not\/}
find that the mass
segregation rate is proportional to the
average mass in the bin (in
contrast to what would be expected for a tracer population of
massive
stars driven by simple dynamical friction; cf.\ Sec.~1.2 and
\citealt{FJPZR02}).  Instead, the following simple expression
provides a good fit (to within a few percent)
to the observed mass-dependence for Model~2: 
\begin{equation} 
\alpha_{\rm ms} t_{\rm rh}(0) = 
\alpha_0 \exp( - m/m_{\rm f}) + \alpha_2, 
\end{equation}
where $m$ is the average stellar mass and the best-fit parameters are
$\alpha_0 = 9.45$, $m_{\rm f} = 21.9\, M_\odot$, and $\alpha_2= -
8.07$.  Note that the mass segregation rate actually approaches a
constant for large $m$ (but of course it is unphysical to extrapolate
beyond $m_{\rm max}$). In the lower panel of
Figure~\ref{fig_mass_segre} we show the average mass within $r_{\rm
h}$, $r_{\rm h}/2$ and $r_{\rm h}/4$. The steady increase of the
average mass in each region is further indication that the mass segregation not only
starts immediately, but also continues until core collapse. For a
smaller number of stars, the average mass within a given radius can
reach saturation \citep[cf.\ Figure~2 of][]{BD98}. Our results for larger $N$
do not show this saturation.

\begin{figure}
\resizebox{\hsize}{!}{\includegraphics[clip]{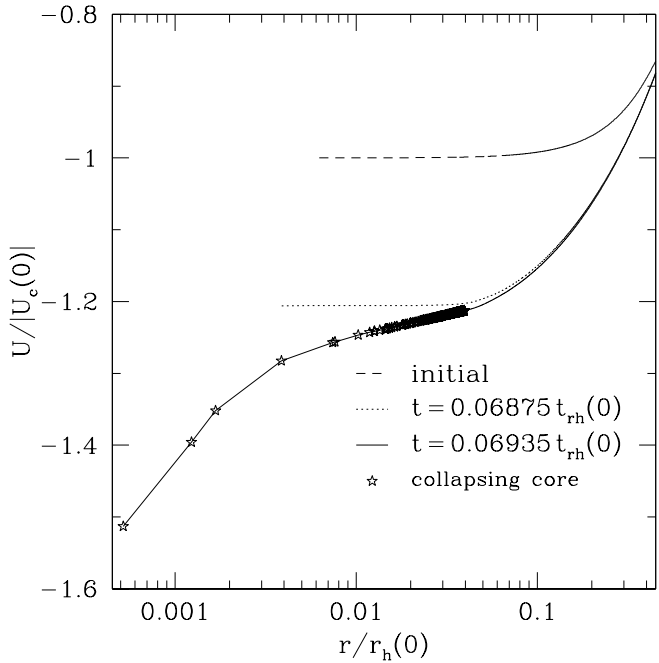}}
\caption{
Evolution of the gravitational potential for Model~2. The potential $U$
is given in units of the absolute value of the initial central potential.
The initial profile is shown
by the dashed line. The profiles just before core collapse and right after
core
collapse are shown by dotted and solid lines, respectively. The positions
of the stars in the collapsing core are marked individually. For this
particular run, the core collapse took place at $t=0.069\, t_{\rm rh}(0)$.
\label{fig_pot_evolution}
}
\end{figure}

Initially and throughout the evolution until core collapse, the cluster
maintains a core-halo structure. However, when the heaviest stars
start dominating the core and the
Spitzer instability occurs, this initial structure is lost.
We demonstrate this behavior in
Figure~\ref{fig_pot_evolution},
where the evolution of the gravitational potential profile is shown.
We have checked that the final structure and, in particular, the formation
of a cusp are not dominated by small number effects
but are instead the result of many massive stars participating in the core
collapse. We illustrate this in Figure~\ref{fig_pot_evolution}
where close to 200 innermost stars,
which constitute 0.2\% of the total mass, are shown explicitly. 

Another important point, which can be deduced from the evolution
of the Lagrange radii, as well as from Figure~\ref{fig_pot_evolution}
(see the lower two curves), is that the outer parts of the
cluster are not much affected by mass segregation and core collapse, except
for the disappearance of the most massive stars.

\begin{figure}
\resizebox{\hsize}{!}{\includegraphics[clip]{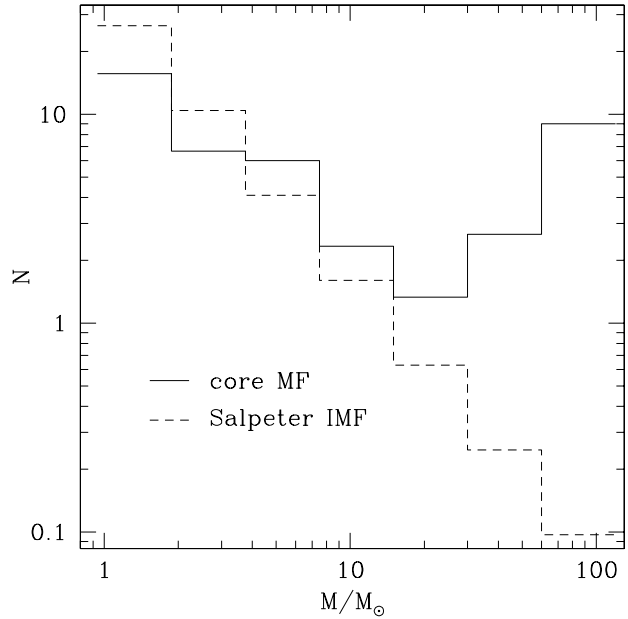}}
\caption{
Comparison between the mass spectrum of stars in the collapsing core (solid
line) and the
Salpeter IMF (dashed line) for Model~2.
Numbers for the core mass function are obtained
by averaging over
three successive snapshots. The Salpeter IMF is normalized so that both
curves account for the same total number of stars.
\label{fig_mf_comp}
}
\end{figure}

As expected, the stars in the collapsing core have a mass distribution much
richer in
heavy stars than the IMF. We compare the mass function at core collapse with
the IMF, for $m\ge1\,M_\odot$, in Figure~\ref{fig_mf_comp}.
The mass function shown in this figure is obtained by choosing the innermost
200 stars at three different times
all before and within $5\times10^{-6}\,t_{\rm rh}(0)$ of core collapse time.

Mass segregation continues until the core is dominated by massive stars.
At this point, the massive stars near the
center dynamically decouple from the rest of the cluster and go into collapse
as a separate subsystem. The stars outside this subsystem act as an energy sink,
and, as a result of their energy gain, expand away from the center of the
cluster. In a single-component system, as the collapse proceeds, the
subset of stars that participate in the collapse becomes smaller and smaller. 
Here instead
the stars that participate in the collapse at the edge of the core
are heavier than the surrounding
stars so they can more efficiently give their energy away and remain together. 
As a result, in systems with a wide mass spectrum and a steep enough IMF, there
is a clear separation between stars in the collapsing core and stars outside the
collapsing core, almost suggesting a real condensation process. The total mass in
the collapsing stars is a crucial property of these systems,
clearly representing an upper limit to the mass of any BH that could
form eventually through the gravitational collapse of a runaway merger remnant. 

To estimate the total mass in the collapsing core,
$M_{\rm cc}$, we
proceed as follows. At every timestep we calculate and record the
Lagrange radii for various mass fractions. For each
Lagrange radius, this provides
typically a few thousand data points per run. The innermost
radii exhibit a great amount of noise for the lowest mass fractions.
We remove this noise in two steps. First, we take arithmetic averages over
60 points and reduce the total number of points to $\sim 50-100$. Then for
each point and using the closest 6 points we fit a
cubic polynomial using least
squares and we evaluate this polynomial at the corresponding time.
This way we both smooth
the Lagrange radii further and estimate their derivatives.
When the system is going into core collapse, a decrease in the collapse rate is
an indication that the results provided by our code are beginning to be
dominated by numerical errors. We stop the simulation when this happens 
and define the core collapse time as the last point before this behavior is
observed.  We then find
the innermost Lagrange radius that has a positive derivative at the time
of core collapse and estimate the mass of the stars that participate in core
collapse as the mass enclosed by this radius.
An investigation of Figure~\ref{fig_model2_zoom}
by eye verifies that this method produces
reasonable results. In practice, we track many more Lagrange radii than
shown in this figure, at intervals of 0.02\% between 0.08\% and 0.36\%.

Because of small-number fluctuations
(and, in particular, the intrinsic noise in the innermost Lagrange radii) 
there is a statistical
uncertainty in all the numbers quoted in Table~\ref{table_models}. This uncertainty is
not unphysical, as real systems will be affected similarly by fluctuations in the
number and specific properties of a relatively small number of massive stars.
To test the robustness of our results and estimate this uncertainty
we have repeated 10--20 simulations with different random seeds for Model~2 and
some of its variants with different numbers of stars (Models~2r, 2s, and~2b).
For these models the numbers in Table~\ref{table_models} are obtained by averaging
the results of the various (physically equivalent) runs. The standard deviations 
obtained for the values of $t_{\rm cc}/t_{\rm rh}(0)$
and $M_{\rm cc}/M_{\rm tot}$ for our typical Model~2 are about 5\% and 20\%, respectively. 
The larger
uncertainty in $M_{\rm cc}/M_{\rm tot}$ is a result of the high sensitivity of this
quantity to noise in the innermost Lagrange radii.

We cannot study the evolution of the cluster past core collapse
without treating in detail the dynamics of the central stars during
collapse, which is beyond the scope of this paper.  In future work we
will include a detailed treatment of stellar collisions in the core
(PaperII, \citealt{FGR03}). Alternatively, one could
use a hybrid method and treat the central part of the cluster with a
direct $N$-body approach \citep{LM85}. However, the simplest
approximation may be to introduce an effective boundary condition
at some very small radius.
\cite{Stodol82}, in his model B2, took this approach, 
but his implementation was not conserving energy.

\subsection{Dependence on the IMF}
The sharp onset of core collapse for our reference model, shown in
Figure~\ref{fig_model2}, is a result of the Spitzer instability.  This
instability is driven by the segregation of the heaviest stars toward
the center and their dynamical decoupling from the rest of the
system. As indicated by the results of Section~\ref{sec_ms_cc}, the
ratio of maximum to average stellar mass in the IMF, $m_{\rm
max}/\left< m \right>$, is an important parameter setting the
timescale for the onset of instability.  There are various ways to
study the dependence of our results on this parameter.  One can use an
IMF different from Salpeter, e.g., Miller-Scalo or Kroupa. In
addition, when using a power-law IMF, changing the slope $\alpha$ or
the maximum mass $m_{\rm max}$ will obviously alter the value of the
ratio $m_{\rm max}/\left< m \right>$.  

\begin{figure}
\resizebox{\hsize}{!}{\includegraphics[clip]{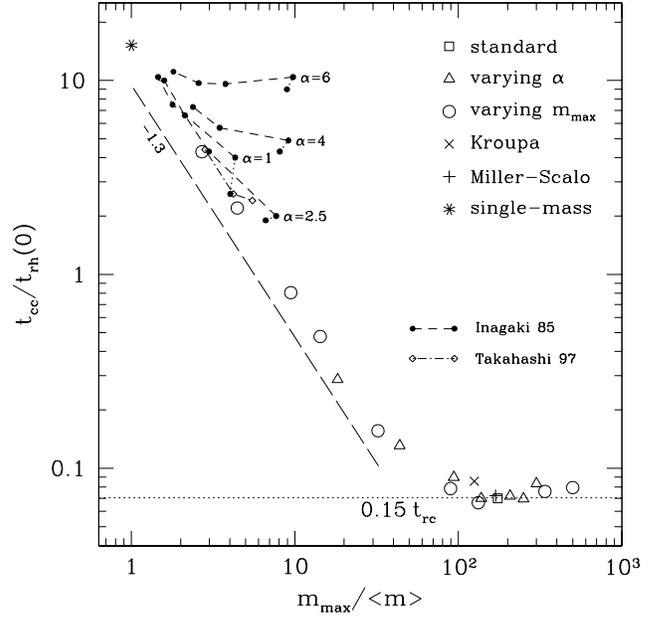}}
\caption{
Dependence of the core collapse time on the shape and width of the
IMF. Initial Plummer models were assumed in all cases.  The horizontal axis shows 
the ratio $m_{\rm max}/\left< m
\right>$ of maximum to mean stellar mass in the IMF.  The square is for our
standard model with a Salpeter IMF, $m_{\rm min}=0.2\, M_\odot$, and
$m_{\rm max}=120\, M_\odot$. The cross and the plus are for the Kroupa
and Miller-Scalo IMF, respectively, with the same limits.  The
triangles are for power-law IMF with various exponents (-1.4,
-1.7, -2.0, -2.2, -2.5, -2.7, -3.0) and the same limits. The circles
are for Salpeter-like IMF with varying upper limits ($m_{\rm max}=\,$ 1, 2,
5, 8, 20, 60, 90, 240, and $360\,M_\odot$).
For comparison, FP results from \citet[][his table 2]{Inagaki85} are 
plotted with small solid dots. Models with the same IMF slope ($\alpha$) 
and $N_{\rm comp}=5$ (number of discrete mass components in the FP simulations) 
are linked together with dashed lines. For each value of $\alpha$, a model with a relatively 
large mass contrast ($m_{\rm max}/m_{\rm min}=10$) was computed with 
$N_{\rm comp}=15$. 
Its core-collapse time is shorter than a 
corresponding model with $N_{\rm comp}=5$ because the mass function is 
better represented. These models are linked to the other points with 
dotted lines.
The three multi-mass models ($N_{\rm comp}=10$) from 
\citet{Takahashi97} are also plotted with small open dots (joined with 
dash-dotted line). They correspond to $\alpha=3.5$, 2.5, and~1.5 (in order 
of increasing $m_{\rm max}/\langle m \rangle$).
\label{fig_imf_comp}
}
\end{figure}

Our results from simulations
for a large number of models exploring these various alternative IMF
are presented in Figure~\ref{fig_imf_comp}. 
Note that for $m_{\rm max}/\langle m \rangle < 40$ our
results suggest a relation $t_{\rm cc} \propto (m_{\rm max}/\langle m
\rangle)^{-1.3}$, also obeyed by all Fokker-Planck models from
\citet{Inagaki85} and \citet{Takahashi97} with $\alpha<3.5$ and a number
of mass components sufficient to ensure proper sampling of the IMF.
However our computations extend to much higher values of $m_{\rm
max}/\langle m \rangle$ than those works. Beyond $m_{\rm max}/\langle
m \rangle\simeq 50$, a domain reached by any realistic IMF, the core
collapse time approaches a constant $\simeq 0.15\, t_{\rm rc}(0)$. Therefore
our main conclusions appear to be {\em independent of the details of the IMF\/}
as long as the number of massive stars in the system is large enough.

For small values of $m_{\rm max}/\left< m \right>$, not only the
timescale but also the very nature of the collapse changes. In these systems
the
evolution timescale, i.e., the relaxation time, of the subsystem that can
decouple from the rest of the cluster is no longer small enough that the
evolution of lighter stars can be neglected. The Lagrange radii for
these models behave similarly to the single-component case, i.e., there is no
clear
separation between collapsing and expanding Lagrange radii. Consequently,
we do not give values of $M_{\rm cc}/M_{\rm tot}$ for these models in
Table~\ref{table_models}.
\cite{Quinlan96}, using FP simulations, also found that, for a
clear decoupling, the relaxation time of the subsystem has to be short
compared to that of the other stars.

\subsection{Dependence on Initial Cluster Concentration}

One expects the core collapse time to depend strongly on the initial density
profile of the cluster, and, in particular, on the central concentration.
A simple and systematic way to examine this dependence is
to use different King models with varying concentration parameter or
$W_0$ \citep[Fig.~4-10]{BT87}.
For single-component clusters this has been done using a variety of methods
\citep{Quinlan96,Einsel96,JNR01,KELSL02}.

\begin{figure*}
\plottwo{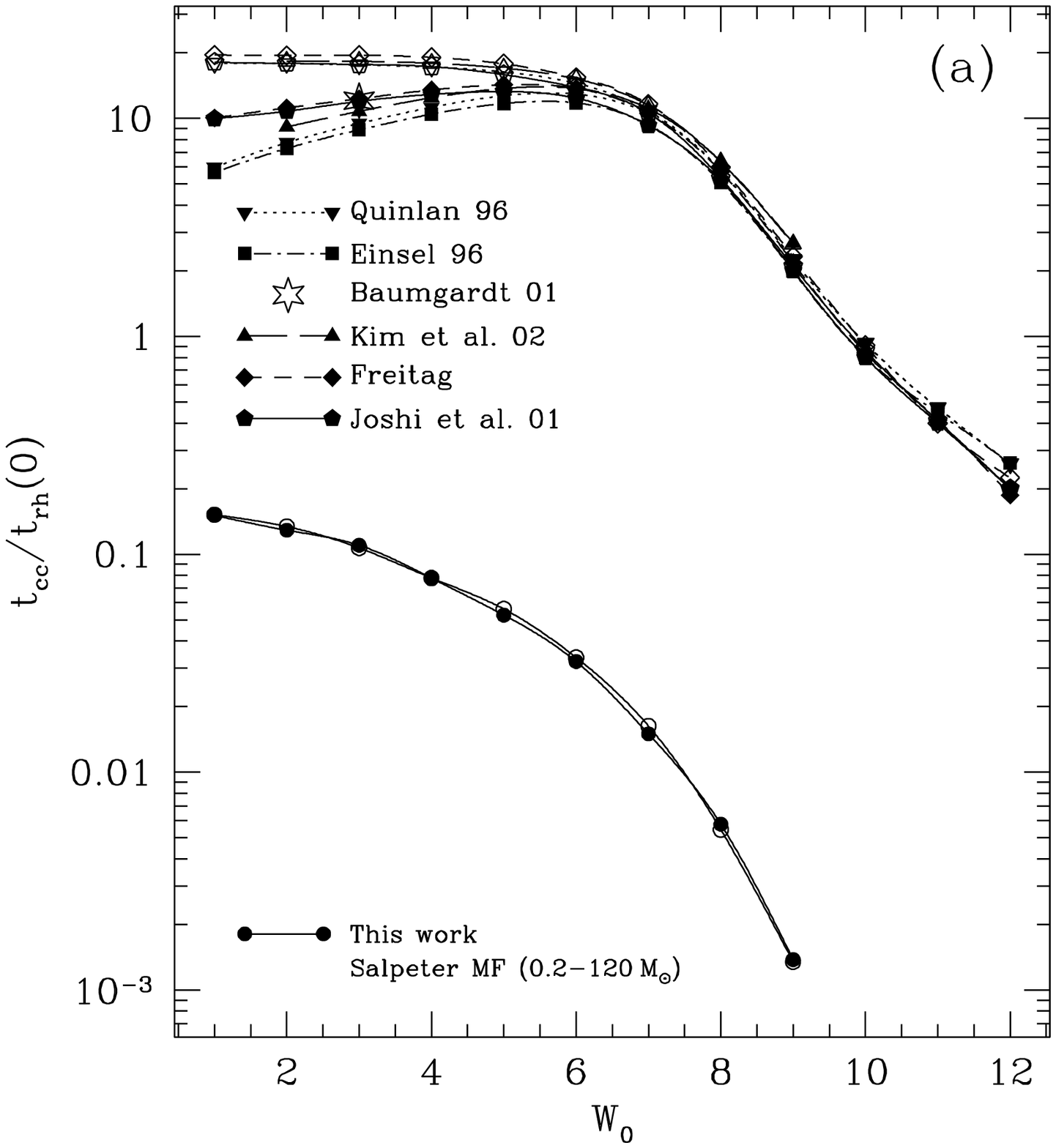}{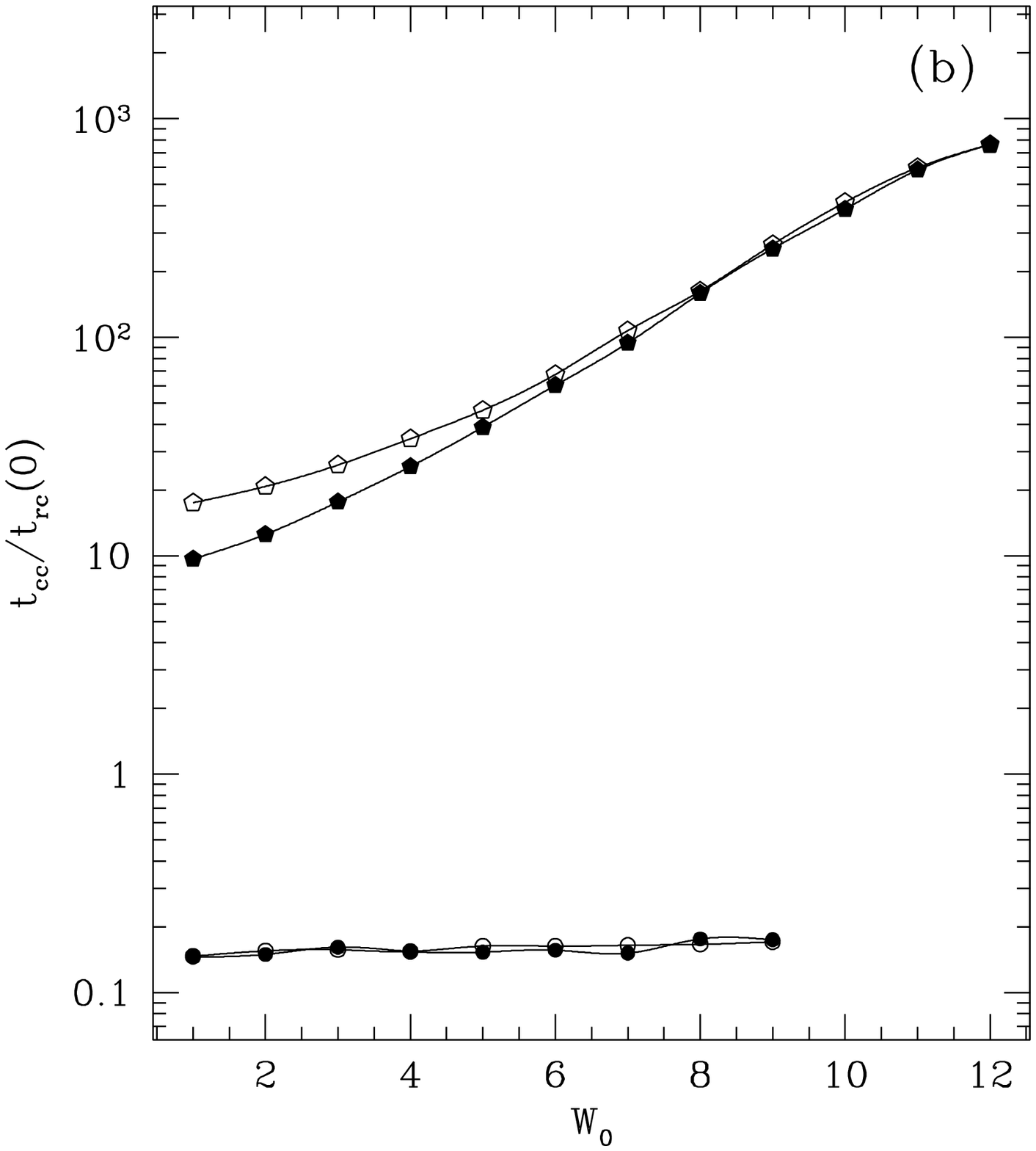}
\caption{
Comparison of core collapse times for various King models with a broad IMF
and with single-mass stars.  (a) The ratio of core
collapse time to half-mass relaxation time. For the single-component
models (top curves), we include results obtained with Fokker-Planck codes
\citep{Quinlan96,Einsel96,KELSL02}, Monte-Carlo codes (\citealt{JNR01};
Freitag, unpublished) and one $W_0=3$ $N$-body simulation by
\citet{Baumgardt01}. Solid and open symbols indicate models with and
without tidal truncation, respectively. (b) The ratio of core collapse time
to {\em central\/} relaxation time. For the sake of clarity, only data from
\citet{JNR01} are plotted for single-component clusters.
\label{fig_tcc_trh}
}
\end{figure*}

We have carried out a number of simulations using King models
with $W_0=1-9$ as initial structure. 
Our simulations include both isolated clusters (with no tidal
boundary enforced, even though the initial models are truncated) and 
clusters with a tidal boundary (assuming a circular orbit in a spherical
galactic potential). This tidal boundary is initially chosen to be at the
tidal radius of the King model and then adjusted as the cluster loses mass
\citep{CW90,JNR01}. 
In Figure~\ref{fig_tcc_trh}a we plot the ratio of core collapse time to 
initial half-mass relaxation time for our simulations, along with results
for single-component clusters. 
The ratio $t_{\rm cc}/t_{\rm rh}(0)$ is much smaller for systems with
mass spectrum, as would be expected. We also see that the core collapse time
for these systems is independent of the presence of a tidal boundary, as expected when
core collapse is driven by local processes within the core
rather than by global energy transfer. This idea is strongly supported by the results in
Figure~\ref{fig_tcc_trh}b, which shows the comparison of the ratio of core
collapse time to {\em central\/} relaxation time, for our simulations and
single-component models. With respect to single-component models, the ratio
$t_{\rm cc}/t_{\rm rc}(0)$ for systems with
a broad mass spectrum shows very little variation.

We have also run two simulations for $\gamma$-models, using
$\gamma=1.0$ (Hernquist model), and $\gamma=1.5$. Initially, models
with $\gamma<2$ have vanishing central velocity dispersion and, for
$\gamma\le 2$, they exhibit a central ``temperature inversion''
\citep{Tetal94}. For single-component clusters, the central region at first
undergoes rapid gravothermal expansion until it becomes isothermal and
``normal'' core collapse can start
\citep{Quinlan96}. Models with $\gamma\le 2$  also have zero initial central 
relaxation time\footnote{Models with $2.0<\gamma<3.0$ have {\em
infinite\/} $t_{\rm rc}(0)$ because their velocity dispersion rises
like $1/r$ near the center \citep{Tetal94}.}. Our results for King
models would then suggest that core collapse induced by central mass
segregation should be extremely fast, if it were not for the opposite
effect of the temperature inversion. For $\gamma=2$, mass segregation
seems to have the upper hand in the competition and it proves
impossible to resolve core collapse because extreme mass segregation
appears nearly instantaneously during the MC runs. Models with
$\gamma=1$ and $\gamma=1.5$ expand at first and then evolve to core
collapse very rapidly. The core collapse time and collapsing core mass
are very hard to determine numerically for these models. In
Table~\ref{table_models}, we give only an upper limit on their core
collapse times. Obviously, these values are so short that their
physical relevance is unclear. It is hard to imagine through which
process such a cluster could be created if one wants it to be
virialized but without initial mass segregation because these
conditions impose constraints on the formation timescale unlikely to
be satisfied in real systems. In addition, note that the local
dynamical time in a $\gamma=1$ model approaches a constant non-zero
value near the center, while, formally, the relaxation time goes to
zero there. This suggests that such models can only provide an
approximate description of real clusters, where finite-number effects
will play a key role near the center, so that the question of
determining the evolution in the limit of very large $N$ is not
well posed. 

\subsection{Initial Mass Segregation}

Naturally we expect that any initial mass segregation in a cluster should lead to
an even shorter core collapse time as the heavier stars are starting
their life closer to the center on average, and therefore do not need
as much time to concentrate there through mass segregation.

\begin{figure}
\resizebox{\hsize}{!}{\includegraphics[clip]{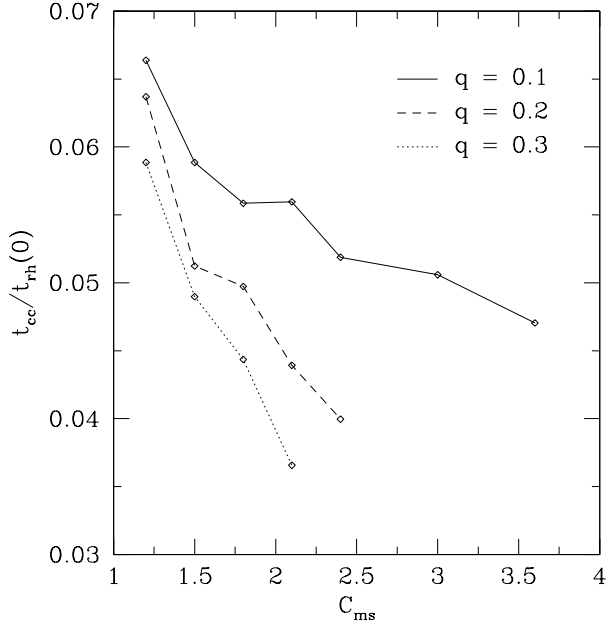}}
\caption{
Dependence of the core collapse time $t_{\rm cc}/t_{\rm rh}(0)$ on initial mass segregation
parameters $C_{\rm ms}$ and $q$ (see Table~\ref{table_ims}).
\label{fig_IMS}
}
\end{figure}

Our results from MC simulations with initial mass segregation (Sec.~3.3) are presented
in Table~\ref{table_ims} and Figure~\ref{fig_IMS}. As expected, we find that
$t_{\rm cc}/t_{\rm rh}(0)$ decreases with increasing $C_{\rm ms}$ (i.e., when stars 
become more and more massive on average in the inner region), and with
increasing $q$ (i.e., when the increase in average stellar mass takes place 
over a larger central region). However, the values
of $M_{\rm cc}/M_{\rm tot}$ do not appear to be affected significantly (within
expected random fluctuations from run to run). The implication for BH formation through
runaway collisions is that
the final BH mass may not be affected by initial mass segregation, while the condition
for runaway growth to occur (see Sec.~5) will be relaxed as the core collapse takes 
place even earlier.

\subsection{Effects of Stellar Evolution}
\label{subsec_stell_evol}

The simulations described so far consider only the
dynamical evolution of the cluster, neglecting the stellar evolution entirely.
We have also carried out a number of simulations to 
understand how stellar evolution and the accompanying mass loss can
modify the dynamical evolution. The stellar evolution treatment we have adopted
is that of \cite{BKB02}, based on the approximations of \cite{HPT00}.

Stellar evolution introduces a new physical clock in the system, 
independent of relaxation. Therefore it is necessary to
specify the physical scale of an initial cluster model (e.g., the half-mass radius
in pc) before starting a simulation so that the relaxation time
$t_{\rm rh}(0)$ can be calculated in years and related to the stellar evolution 
timescale. The only other parameter to be specified
is the metallicity $Z$, which plays an important role in the calculation of
stellar evolution mass loss. In our treatment of stellar evolution, all wind mass
loss rates are proportional to $Z^{1/2}$. Since stellar evolution introduces two
new parameters in our initial models, a full exploration of the initial parameter space
is clearly impossible. Fortunately, we will see that the effects of mass loss
on the evolution to core collapse are rather unimportant, and so a systematic 
study is unnecessary at this point.

\begin{figure}
\resizebox{\hsize}{!}{\includegraphics[clip]{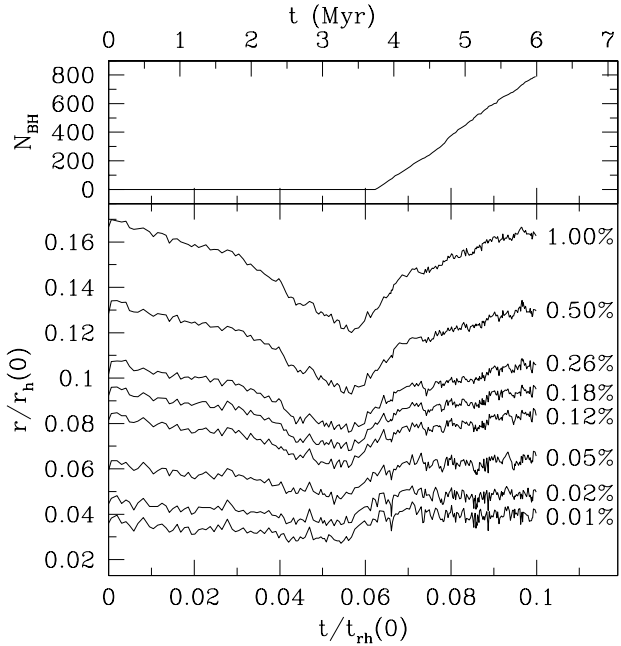}}
\caption{Prevention of core collapse due to stellar evolution mass loss. 
The lower panel shows the evolution of the Lagrange radii.  The upper panel gives the number
of black holes in the cluster (formed by evolving massive stars). The time is given in units of $t_{\rm
rh}(0)$ for the bottom and middle axes, and in Myr along the top
axis.  The first massive stars leave the main sequence around $3\,$Myr.
\label{fig_mass_loss}
}
\end{figure}

We first illustrate how {\em post-main-sequence\/} evolution of massive stars
can prevent core
collapse. In Figure~\ref{fig_mass_loss} we show the evolution of a system
similar to Model~2 with $t_{\rm rh}(0)=60\,{\rm Myr}$ and
$Z=10^{-4}$. At such low metallicity, there is little mass loss on the
main sequence, so one expects $t_{\rm cc}\simeq 0.07\times 60\,{\rm Myr} = 
4.2\,{\rm Myr}$ without evolution beyond the main sequence. However, massive stars 
evolve off the main sequence after about $3\,$Myr, before
core collapse has occurred. Once they evolve off the main sequence, they typically lose up to
$2/3$ of their mass rapidly.  At the end of their life, which
extends about 10\,\% beyond their main-sequence lifetime, these stars undergo
supernova explosions
and their cores collapse to BH. In the upper panel of Figure~\ref{fig_mass_loss}
we plot the number of these BH in the cluster.  The significant mass loss during
the late stages of stellar evolution causes the cluster core to expand so that
collapse is prevented and the system then goes into long-term dynamical
evolution. 

Varying $Z$ from $10^{-4}$ to 0.02 (the solar value), we found that this reversal of
core contraction occurs for any metallicity when massive stars are allowed to
evolve off the main sequence. Higher metallicities of course lead to increased mass 
loss and an even stronger tendency for the cluster core to expand.
We have not considered the case of $Z=0$, i.e., a cluster of Pop~III stars. 
These stars may collapse to BH that
incorporate essentially the entire initial stellar mass \citep{HWFL03}. The
dynamical evolution of the cluster would then be unaffected by mass loss.
Whenever a runaway is avoided, a dense cluster of relatively massive primordial 
BH would then form near the center. This is
an intriguing possibility to keep in mind for future consideration,
although the IMF of Pop~III stars is essentially unknown (but see,
e.g., \citealt{NU01,AbiaEtAl01}) and recent hydrodynamic simulations
suggest that the first stars may actually form isolated rather
than in clusters \citep{ABN02}.

We have also examined the possibility that mass loss {\em on the main sequence\/} may
already be a source of indirect heating strong enough to reverse
core collapse or, at least, delay it significantly. Our results indicate
that wind mass loss from main-sequence stars alone can never prevent core collapse and 
that the delay introduced remains very small, even for high metallicities. It is of
course always possible to fine-tune the relaxation time so that the
mass loss slows down the evolution just enough to allow a few massive
stars to evolve off the main sequence and stop contraction. But such models
can only represent a very small
domain in the parameter space of initial conditions. For example, 
in a system like Model~2 with $t_{\rm rh}(0)=40\,{\rm Myr}$, the core collapse
time increases only by about 5\% for $Z=0.02$. For $t_{\rm rh}(0)=50\,{\rm Myr}$, the
evolution is just delayed long enough to prevent core collapse for $Z=0.02$,
but not for $Z=0.001$. One has to increase the relaxation time to 
$t_{\rm rh}(0)=60\,{\rm Myr}$ for stellar evolution to prevent core collapse for all
metallicities.

In conclusion, we find that post-main-sequence mass loss can be strong enough to
prevent early core collapse, but that it does not significantly tighten
the condition 
(on $t_{\rm rh}(0)$) for core collapse to occur. Mass loss from main-sequence stars always plays
 a negligible role.

\section{Summary and Discussion}
\label{sec_disc}

It has long been realized that a mass spectrum accelerates the
dynamical evolution of dense star clusters through the process of mass
segregation \citep{Aarseth66,Henon71,Wielen75}. This is a consequence
of the statistical tendency of two-body gravitational encounters to
establish energy equipartition between stars of different masses. For
any realistic IMF, equipartition can never be achieved because the
massive stars quickly form a separate self-gravitating subcluster (a system with
negative effective heat capacity) as they segregate near the center
by transferring kinetic energy to lighter stars
\citep{Spitzer69,Vishniac78,IW84,IS85}. This instability leads to a 
rapid core collapse: after a finite time
$t_{\rm cc}$, the central density of stars would actually become
infinite in the absence of finite size effects (both physical collisions
between stars and the finite number of stars in the core).

With the exception of a few recent $N$-body simulations
\citep{PZMMH99,PZM02}, all previous investigations of
core collapse in clusters with a mass spectrum have considered a
relatively narrow range of stellar masses ($m_{\rm max}/m_{\rm min} \ll
100$). This restriction is appropriate for old globular clusters in
which stellar evolution probably had time to remove all massive
stars before significant relaxation took place. In
contrast, in the present study, we ignored stellar evolution to
concentrate on systems in which, by assumption, core collapse occurs
before even the most massive stars (with $m\sim 100\,M_\odot$) leave
the MS. 

Through a set of high-resolution Monte Carlo simulations of clusters
with a variety of IMF and structural parameters (concentration,
presence or absence of tidal truncation or initial mass segregation),
we established two important results. The first concerns the time
needed to go deep into core collapse ($t_{\rm cc}$). As the evolution to core collapse
is driven by relaxation processes, $t_{\rm cc}$ can always be written
as proportional to the initial half-mass or
central relaxation time ($t_{\rm rh}(0)$, $t_{\rm rc}(0)$). The proportionality constant
does not depend on the size of the cluster or the number of
stars, as long as they are numerous enough to avoid small-number
effects. For isolated clusters with an IMF of realistic
slope ($\alpha\simeq 2.5$ in the high-mass range), we find 
\begin{equation}
t_{\rm cc}/t_{\rm rh}(0) \propto \mu^{-\delta}\ \ \mbox{with}\ \ \mu=\frac{m_{\rm max}}{\langle m
\rangle}\ \ \mbox{and}\ \ \delta \simeq 1.3,
\end{equation}
as $\mu$ increases from 1 (single-mass) to $\sim 50$. This result
extends and agrees nicely with previous work based on FP simulations
\citep{Inagaki85,Takahashi97}. From simple arguments about
dynamical friction, one would naively expect a linear relation with
$\delta=1$ \citep{BT87}. 
The steeper dependence found in our numerical results may be related to
the fact that the instability causing the core collapse is
triggered only after some critical number, $N_{\rm cr}$, of sufficiently
massive stars have drifted to the core by dynamical friction. If one
has to go out to radius $R_{\rm cr}$ to find this number of heavy
stars, the required timescale will be of order $\mu^{-1}t_{\rm
rel}(R_{\rm cr})$. For systems with increasing $\mu$, it is reasonable
to expect $N_{\rm cr}$ and, consequently, $R_{\rm cr}$ and $t_{\rm
rel}(R_{\rm cr})$ to decrease. While this provides a plausible
explanation, further investigations will be needed to confirm it, 
in particular studies making use of the gaseous model
of cluster dynamics \citep{FASS03}.

For $\mu > 50$, the core collapse time saturates to 
\begin{equation}t_{\rm cc} \sim
0.15\,t_{\rm rc}(0),
\end{equation}
a key result given that any realistic IMF is
likely to have $\mu > 100$. This value of $t_{\rm cc}$ is at least two
orders of magnitude shorter than the core collapse time for
single-component clusters. Moreover, we find this
$t_{\rm cc}/t_{\rm rc}(0)$ ratio to hold for all the models we have
considered that have a finite $t_{\rm rc}(0)$. This includes King
models (with or without tidal truncation) with $W_0$ ranging from 1 to
9, a sequence along which $t_{\rm cc}/t_{\rm rh}(0)$ decreases by more
than $\sim100$. Our result that the core collapse time is set
fundamentally by $t_{\rm rc}(0)$ rather than $t_{\rm rh}(0)$ reflects
an important difference between single-component clusters and systems
with a broad mass spectrum: in systems with a mass spectrum, core
collapse is driven ultimately by energy transfer occurring {\em
locally\/} in the core; in
contrast, relaxation in a single-component system is a {\em global\/}
process taking place on all scales, resulting in a timescale
comparable to $t_{\rm rh}$
\citep{Inagaki85}. For Plummer models or King models with moderate 
concentration ($W_0\le 4$), we find $t_{\rm cc}\simeq
0.07-0.15\,t_{\rm rh}(0)$, in good agreement with the results of $N$-body
simulations by \citet{PZM02}\footnote{Although the fundamental relation is
 between $t_{\rm cc}$ and $t_{\rm rc}(0)$, $t_{\rm rh}$ is, by far, easier to
estimate observationally.}. 

The main-sequence lifetime $t_*$ of very massive stars approaches
a constant value, of about $3-4\,$Myr, with only weak dependence on
metallicity, rotation, or the mass $m$ of the star, as long as
$m \ga 50\,M_\odot$ \citep{SSMM92,MM00,MM01}. It exhibits little 
variation with $m$ because such massive stars are
nearly Eddington-limited, hence $L \propto m$, where $L$ is the
luminosity, and $t_* \propto f_{\rm c} m / L
\simeq\,$constant if the fractional mass of the convective core, 
$f_{\rm c}$, does not depend too strongly on the stellar mass\footnote{If
the mass segregation time scale $t_{\rm s}\propto 1/m$ is compared to
the ``usual'' relation $t_*\propto 1/m^3$, one could conclude erroneously
that massive stars never play an important role in core collapse
\citep{Applegate86}. However, the approximate $1/m^3$ scaling of the stellar 
lifetime applies only for $m\lesssim 10\,M_\odot$.}.
Therefore, the necessary condition for core
collapse not being stopped by stellar evolution can be written
\begin{equation}
t_{\rm rh}(0)\lesssim 30\,\mathrm{Myr}.
\end{equation}
Interestingly, we also find that core collapse cannot be
prevented by mass loss through stellar winds on the MS. Such mass loss
can only delay the evolution to core collapse very slightly. Even for
high metallicities and hence strong winds, only clusters for which the
core collapse time would otherwise be very close to
the critical value (i.e., just below $t_*$) can see their fate changed through
wind mass loss if $t_{\rm cc}$ increases to become slightly longer than 
$t_*$. These clusters must represent a very small domain in the parameter space
of initial conditions.

Our second important finding is that, as the collapse proceeds, the mass
contained in the ever-shrinking core converges to a non-zero value,
\begin{equation}
M_{\rm cc} \simeq 0.001-0.002\,M_{\rm tot},
\end{equation}
which depends only weakly
on the properties of the initial cluster. This contrasts
again with single-component clusters which, exhibiting self-similar
collapse, have $M_{\rm cc}\rightarrow 0$ as $t \rightarrow t_{\rm
cc}$.

This value of $M_{\rm cc}/M_{\rm tot}$ is in remarkable agreement with
the fractional mass of a dark object, presumably a massive BH,
discovered at the center of galaxies (in which case, $M_{\rm tot}$
stands for the mass in the spheroidal component) and, possibly,
globular clusters like M15 and G1. This suggests that the collapsing
core may well be the progenitor of a central IMBH. However, for
a galactic nucleus, the requirement that $t_{\rm cc}\simeq 0.15\,t_{\rm
rc}(0) < 3-4\,$Myr would imply that, either it evolved from a cluster
with short (central) relaxation time to its present-day state with $t_{\rm rh}\ge 10^8\,$yr  
\citep{LFAGS98}, or that the growth of the central BH was predominantly from accretion of
smaller star clusters with short relaxation times (each one carrying an
IMBH at its center; \citealt{Ebisuzakietal01,HM03}). M15 and G1 also have
long half-mass relaxation times ($\sim1\,$Gyr for M15,
\citealt{DullEtAl97}, and $30-50\,$Gyr for G1,
\citealt{MeylanEtAl01,Betal03b}) indicating that the process we 
envision could only have taken place if these clusters were born with
much more centrally concentrated density profiles than observed today\footnote{It is not clear that the 
present-day
value of $t_{\rm rh}$ for a cluster, which can be estimated from observations, must be
close to $t_{\rm rh}(0)$.} (for M15, an {\em initial\/} value of
$W_0\gtrsim 8$ seems required; see Fig.~\ref{fig_tcc_trh}). However,
there is no doubt that young star clusters with half-mass relaxation times
shorter than the critical value of 30\,Myr do exist. The Arches cluster near
the center of our Galaxy is the best established example, with $t_{\rm
rh}$ probably shorter than 10\,Myr \citep{FKMSRM99}, while the
Quintuplet cluster appears to lie very close to the critical value
\citep{FMcLM99}. It is also possible that some super 
star clusters, the birthplaces of most stars in starburst environments,
have sufficiently short relaxation times \cite[e.g.][]{HF96}.

Although the final fate of the stars that participate in core collapse is not
completely certain, very high rates of physical collisions are expected in the core. 
Primordial binaries are probably unable to stop the collapse and prevent these collisions.
In smaller systems like globular clusters interactions with hard binaries
 may in fact increase the collision rate.
In more extreme environments like (proto-)galactic nuclei, the velocity dispersion
is so high ($\sim10^2-10^3\,{\rm km}\,{\rm s}^{-1}$) that most primordial binaries
are soft and will have been disrupted before they can play a role in the collapse.
The main uncertainty
is then whether these collisions, occurring at relatively high
velocity, allow a massive star to gain mass in a runaway fashion, as
suggested by analytical models (based on the dependence of the cross section on
the mass and neglecting collisional mass-loss; \citealt[][and references therein]{MG02}), 
or, on the contrary, grind it down progressively. Our preliminary results from 
MC simulations including a realistic treatment
of stellar collisions \citep{RFG03} indicate that the formation of
a very massive star through runaway collisions is a likely outcome. 
More extensive calculations, as well as a detailed discussion of the likely fate of the
merger remnant, will be presented in Paper~II.

\acknowledgements
We thank Holger Baumgardt for discussions and for providing data from
his $N$-body simulations. We are grateful to Rainer Spurzem for making
the code for gaseous models ({\it SPEDI}) available and for his
assistance in using it. Piero Spinnato and Simon Portegies Zwart kindly
gave us the initial condition data for the ``dynamical friction'' run
presented in the Appendix. We thank K.\ Belczynski for providing us
the population synthesis code {\it StarTrack\/} and for his help in
integrating it into our MC code, and to
John Fregeau for useful discussions.
This work was supported by NASA ATP
Grant NAG5-12044 and NSF Grant AST-0206276 to Northwestern University.
The work of MF is funded by the Sonder\-forschungs\-bereich (SFB) 439
`Galaxies in the Young Universe' (subproject A5) of the German Science
Foundation (DFG) at the University of Heidelberg. MF thanks the
Theoretical Astrophysics Group at Northwestern University for
financial support and hospitality during several visits which made
this collaboration possible.

\appendix
\section{Applicability of the Monte Carlo Method}
The Monte Carlo method has been applied previously to two-component
systems \citep[see, e.g.,][]{WJR00} and to systems with a
continuous mass function covering a relatively narrow range (with
$m_{\rm max}/m_{\rm min}\lesssim30$; see the references in
Section~\ref{sec_MC_sim}). However, one may question the applicability
of the method to systems with a much broader mass spectrum. In
particular, the nature of the energy transfer in these systems is
different than in systems with small $m_{\rm max}/m_{\rm min}$ ratio
(See the discussion in Sec.~1.2).  In addition, the gravitational
potential is assumed to change smoothly within the cluster in the MC
method. However, in a system containing a broad IMF, a relatively
small number of very massive stars could introduce significant
discrete changes in the slope of the
potential. As the positions of these steps change randomly from one
iteration to the next, one may worry that this could introduce a
significant amount of spurious relaxation in the system.  
In this Appendix, we provide numerical tests and
theoretical arguments showing that the MC method is indeed applicable
to systems with a very broad mass spectrum, such as those under
consideration in this paper.

\subsection{Spurious Relaxation}

In MC simulations, the orbits of the stars are calculated using a potential
assumed to be spherical and smooth. This potential is constructed numerically from a
radial distribution of point masses and therefore includes random
fluctuations that can lead to spurious relaxation effects \citep{Henon71}.
To test the importance of these effects, we carried out simulations in which
we turned off physical relaxation (by skipping the perturbation of stars).
Ideally, in this case, the cluster should maintain a constant state indefinitely and
none of its properties should change over time. In practice, however, it is
sufficient to make sure that the changes are not significant compared to changes
occurring in the presence of physical relaxation. 

\begin{figure}
\resizebox{0.57\hsize}{!}{\includegraphics[clip]{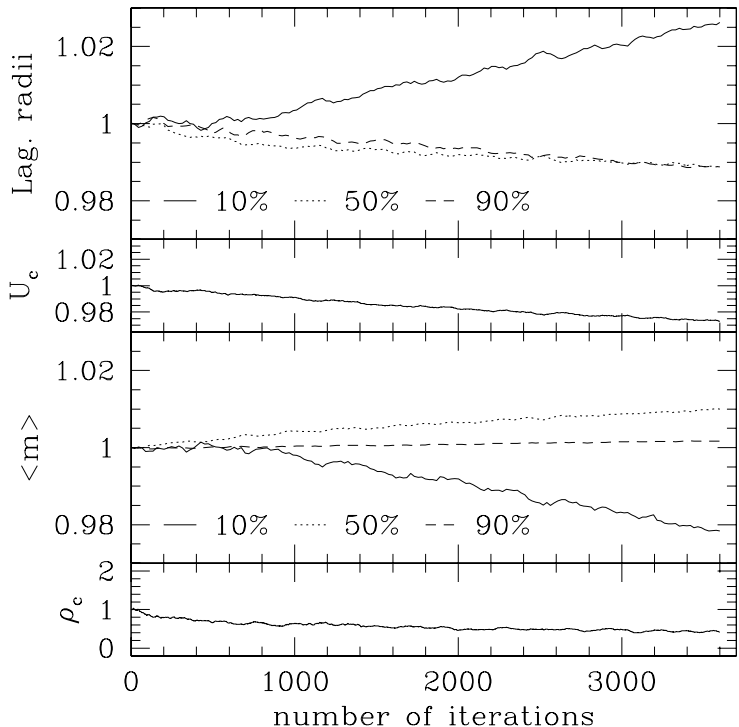}}
\hfill
\parbox[b]{0.4\hsize}{\caption{Test of spurious relaxation in Model~2. Here we show the evolution of 
the 10\%, 50\%, and 90\%
Lagrange radii, the central potential, the average mass within Lagrange
radii, and the central density. All quantities are divided by
the average of their first ten values in the run.
\label{fig_spur_relax}}}
\end{figure}

In one of our tests, we let a MC
simulation, for our typical Model~2, run for 3600 iteration steps, which is
twice the number of steps normally required for this model to reach core collapse. We plot
various relevant quantities showing the evolution of the system in
Figure~\ref{fig_spur_relax}. It is clear that the deviations in these quantities are much
smaller than the changes due to physical relaxation, so we conclude that
the effects of spurious relaxation is not significant.

\subsection{Relaxation and Dynamical Friction}

The energy transfer from massive to lighter stars and the segregation of massive
stars to the center of the cluster are processes similar to simple
dynamical friction for a heavy test particle embedded in a background of lighter 
stars \citep[Sec.~7.1]{BT87}.
Some of the processes  that govern the dynamics of dense stellar systems cannot
be treated simply as part of two-body relaxation. Well-known examples include
collisions, binary interactions, and three-body formation of binaries. A natural
question to ask is whether dynamical friction is a process that can be modeled
correctly by our implementation of two-body relaxation in the MC method. The answer 
is central to the applicability of the MC method to our problem.

The frictional drag on a massive object is proportional to the {\em mass density\/}
of light background stars, and is independent of the masses of individual
stars \citep[see][eq.~7-18]{BT87}. So the effect of dynamical friction
would be unaltered, for example, if every light star were replaced by two
stars of half the mass. However, the treatment of relaxation in the MC method (Sec.\ 2.1) 
does depend on the mass ratio of the two particles considered 
and would be altered by such a replacement. The ratio
of relaxation to crossing time is given by
\begin{equation}
\frac{t_{\rm r}}{t_{\rm dyn}} \propto \frac{N}{\ln N}
\end{equation}
which would evidently change if the stars in a system were replaced by many more
stars with smaller masses. This might suggest that
the MC method cannot follow dynamical friction correctly.
However, note that the relaxation time can be written \citep[eq.~3-37]{Spitzer87}
\begin{equation}
t_{\rm r} \propto \frac{v^3}{n\, m^2}
\end{equation}
and the timescale for dynamical friction is proportional to 
\citep[eq.~7-18]{BT87}
\begin{equation}
t_{\rm DF} \propto \frac{v_M^3}{n\, m\, M} .
\end{equation}
Here, $v$ is the average velocity of background stars, of mass $m$ and number density $n$, 
and $v_M$ is the velocity of the massive test particle, of mass $M$.
When $v_M$ is comparable to $v$ (which is true, for example, if the massive particle is 
on a circular orbit; see below), we conclude that
\begin{equation}
t_{\rm DF} \propto \frac{m}{M} t_{\rm r} \propto \frac{v^3}{\rho M}.
\end{equation}
The first proportionality implies that dynamical friction can be treated as a
relaxation process. In the second, we show explicitly the dependence on the
total mass density $\rho$ of the background stars.
This result is also established by several numerical calculations 
\citep{SS75,BD98,FJPZR02}.

\begin{figure}
\resizebox{0.57\hsize}{!}{\includegraphics[clip]{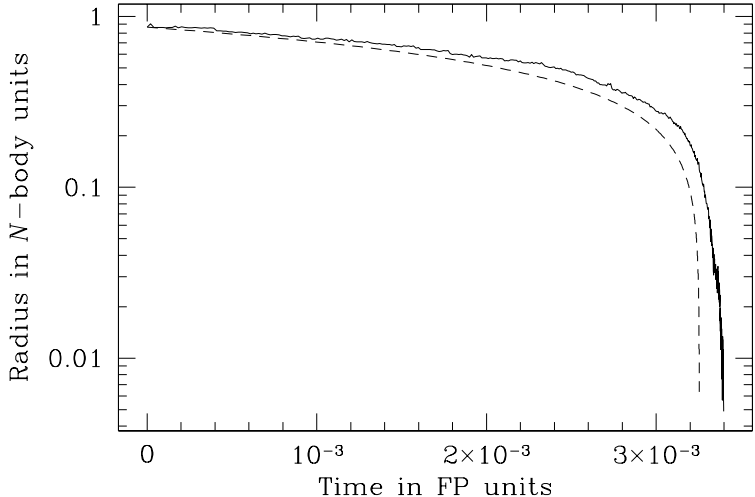}}
\hfill
\parbox[b]{0.4\hsize}{\caption{
Orbital decay of a massive particle through dynamical friction. We plot
the distance from the center, in $N$-body units, as a function of time in 
Fokker-Planck units (Sec.\ 3.4). In these
units, the evolution does not depend on the value of $\gamma_{\rm
c}$. We use our MC code to follow the decay of one massive object in
a cluster of much lighter stars. The initial conditions are identical
to those used by
\citet{SFPZ02}. The solid line shows the MC simulation result while the 
dashed line is the analytic solution from the standard
Chandrasekhar theory of dynamical friction. See text for details.
\label{fig_dyn_fric}}}
\end{figure}

To demonstrate that the MC algorithm is indeed able to handle simple
dynamical friction, we have simulated a system consisting of a single
massive object (mass $m_2$) in a cluster of much lighter particles
(mass $m_1$, number $N_1$). The initial conditions are identical to
those used for the $N_1=400\,000$ model of \citet{SFPZ02}, with
$m_2/m_1=211$, hence the total number of particles used for this test
is 400\,001. \citeauthor{SFPZ02} were interested in the spiral-in of
an IMBH in a galactic nucleus. The massive particle is initially on a
circular orbit at a relatively large distance from the center,
$r(0)$. For the power-law density profile used to represent the
cluster, $\rho(r)\propto r^{-\gamma}$, one can predict analytically
the decay $R(t)$ using Chandrasekhar's formula for dynamical friction
\citep[][eq.~5]{SFPZ02}. In Figure~\ref{fig_dyn_fric}, we compare 
the spiral-in of the massive particle with the analytical solution. In
the formula, we have set $\gamma=2.3$ to fit (visually) the
density profile of the cluster. The initial $N$-body set-up was provided by
\citeauthor{SFPZ02} and converted directly into MC particles. The agreement 
between the MC run and the analytical formula is very satisfactory, especially
considering that the density profile does not follow the exact
power law and in fact evolves slightly during the simulation as a result of
relaxation between light particles. Note that this good agreement
does not validate the assumptions used in deriving Chandrasekhar's
formula (e.g., neglecting large-angle scatterings, correlations between
light particles, and gradients in the properties of the cluster)
because the MC method relies basically on the same set of assumptions. 
In particular, $\gamma_{\rm c}$ is a free parameter in both approaches. What this result
demonstrates is that these assumptions are correctly implemented in our MC
code.

\bibliographystyle{apj}

\clearpage
\begin{thebibliography}{125}
\expandafter\ifx\csname natexlab\endcsname\relax\def\natexlab#1{#1}\fi

\bibitem[{{Aarseth}(1966)}]{Aarseth66}
{Aarseth}, S.~J. 1966, \mnras, 132, 35

\bibitem[{{Aarseth} {et~al.}(1974){Aarseth}, {H{\' e}non}, \& {Wielen}}]{AHW74}
{Aarseth}, S.~J., {H{\' e}non}, M., \& {Wielen}, R. 1974, \aap, 37, 183

\bibitem[{{Abel} {et~al.}(2002){Abel}, {Bryan}, \& {Norman}}]{ABN02}
{Abel}, T., {Bryan}, G.~L., \& {Norman}, M.~L. 2002, Science, 295, 93

\bibitem[{{Abia} {et~al.}(2001){Abia}, {Dom{\'{\i}}nguez}, {Straniero},
  {Limongi}, {Chieffi}, \& {Isern}}]{AbiaEtAl01}
{Abia}, C., {Dom{\'{\i}}nguez}, I., {Straniero}, O., {Limongi}, M., {Chieffi},
  A., \& {Isern}, J. 2001, \apj, 557, 126

\bibitem[{{Applegate}(1986)}]{Applegate86}
{Applegate}, J.~H. 1986, \apj, 301, 132

\bibitem[{{Bacon} {et~al.}(1996){Bacon}, {Sigurdsson}, \& {Davies}}]{BSD96}
{Bacon}, D., {Sigurdsson}, S., \& {Davies}, M.~B. 1996, \mnras, 281, 830

\bibitem[{{Baumgardt}(2001)}]{Baumgardt01}
{Baumgardt}, H. 2001, \mnras, 325, 1323

\bibitem[{{Baumgardt} {et~al.}(2003{\natexlab{a}}){Baumgardt}, {Hut}, {Makino},
  {McMillan}, \& {Portegies Zwart}}]{Betal03}
{Baumgardt}, H., {Hut}, P., {Makino}, J., {McMillan}, S., \& {Portegies Zwart},
  S. 2003{\natexlab{a}}, \apjl, 582, L21

\bibitem[{{Baumgardt} {et~al.}(2003{\natexlab{b}}){Baumgardt}, {Makino}, {Hut},
  {McMillan}, \& {Portegies Zwart}}]{Betal03b}
{Baumgardt}, H., {Makino}, J., {Hut}, P., {McMillan}, S., \& {Portegies Zwart},
  S. 2003{\natexlab{b}}, \apjl, 589, L25

\bibitem[{{Belczynski} {et~al.}(2002){Belczynski}, {Kalogera}, \&
  {Bulik}}]{BKB02}
{Belczynski}, K., {Kalogera}, V., \& {Bulik}, T. 2002, \apj, 572, 407

\bibitem[{{Binney} \& {Tremaine}(1987)}]{BT87}
{Binney}, J., \& {Tremaine}, S. 1987, {Galactic dynamics} (Princeton, NJ,
  Princeton University Press)

\bibitem[{{Bonnell} \& {Bate}(2002)}]{BB02}
{Bonnell}, I.~A., \& {Bate}, M.~R. 2002, \mnras, 336, 659

\bibitem[{{Bonnell} {et~al.}(2001){Bonnell}, {Bate}, {Clarke}, \&
  {Pringle}}]{BBCP01}
{Bonnell}, I.~A., {Bate}, M.~R., {Clarke}, C.~J., \& {Pringle}, J.~E. 2001,
  \mnras, 323, 785

\bibitem[{{Bonnell} \& {Davies}(1998)}]{BD98}
{Bonnell}, I.~A., \& {Davies}, M.~B. 1998, \mnras, 295, 691

\bibitem[{{Chernoff} \& {Huang}(1996)}]{CH96}
{Chernoff}, D.~F., \& {Huang}, X. 1996, in IAU Symp. 174: Dynamical Evolution
  of Star Clusters: Confrontation of Theory and Observations, 263

\bibitem[{{Chernoff} \& {Weinberg}(1990)}]{CW90}
{Chernoff}, D.~F., \& {Weinberg}, M.~D. 1990, \apj, 351, 121

\bibitem[{{Clarke}(2003)}]{Clarke03}
{Clarke}, C.~J. 2003, in Carnegie Observatories Astrophysics Series, Vol. 1:
  Coevolution of Black Holes and Galaxies

\bibitem[{{Cutler} \& {Thorne}(2002)}]{CT02}
{Cutler}, C., \& {Thorne}, K.~S. 2002, in Proceedings of GR16, Durban, South
  Africa, gr-qc/0204090

\bibitem[{{de Grijs} {et~al.}(2003){de Grijs}, {Gilmore}, \&
  {Johnson}}]{dGGJ03}
{de Grijs}, R., {Gilmore}, G.~F., \& {Johnson}, R.~A. 2003, in The Local Group
  as an Astrophysical Laboratory, STScI Symp., May 2003, Baltimore (USA), ed.
  Livio M, astro-ph/0305262

\bibitem[{{Dehnen}(1993)}]{Dehnen93}
{Dehnen}, W. 1993, \mnras, 265, 250

\bibitem[{{Dull} {et~al.}(1997){Dull}, {Cohn}, {Lugger}, {Murphy}, {Seitzer},
  {Callanan}, {Rutten}, \& {Charles}}]{DullEtAl97}
{Dull}, J.~D., {Cohn}, H.~N., {Lugger}, P.~M., {Murphy}, B.~W., {Seitzer},
  P.~O., {Callanan}, P.~J., {Rutten}, R.~G.~M., \& {Charles}, P.~A. 1997, \apj,
  481, 267

%
\bibitem[{{Ebisuzaki} {et~al.}(2001)}]{Ebisuzakietal01}
{Ebisuzaki}, T., et al. 2001, \apjl, 562, L19

\bibitem[{{Eckart} {et~al.}(2002){Eckart}, {Genzel}, {Ott}, \& {Sch{\"
  o}del}}]{EckartEtAl02}
{Eckart}, A., {Genzel}, R., {Ott}, T., \& {Sch{\" o}del}, R. 2002, \mnras, 331,
  917

\bibitem[{{Eggleton} {et~al.}(1989){Eggleton}, {Tout}, \& {Fitchett}}]{ETF89}
{Eggleton}, P.~P., {Tout}, C.~A., \& {Fitchett}, M.~J. 1989, \apj, 347, 998

\bibitem[{{Einsel}(1996)}]{Einsel96}
{Einsel}, M. 1996, PhD thesis, Christian-Albrechts-Universit{\"{a}}t zu Kiel

\bibitem[{{Elson} {et~al.}(1987){Elson}, {Hut}, \& {Inagaki}}]{EHI87}
{Elson}, R., {Hut}, P., \& {Inagaki}, S. 1987, \araa, 25, 565

\bibitem[Farouki \& Salpeter(1994)]{FS94} Farouki, R.~T.~\& 
Salpeter, E.~E.\ 1994, \apj, 427, 676 

\bibitem[{{Ferrarese}(2002)}]{Ferrarese02}
{Ferrarese}, L. 2002, \apj, 578, 90

\bibitem[{{Ferrarese} \& {Merritt}(2000)}]{FM00}
{Ferrarese}, L., \& {Merritt}, D. 2000, \apjl, 539, L9

\bibitem[{{Ferrarese} {et~al.}(2001){Ferrarese}, {Pogge}, {Peterson},
  {Merritt}, {Wandel}, \& {Joseph}}]{FerrareseEtAl01}
{Ferrarese}, L., {Pogge}, R.~W., {Peterson}, B.~M., {Merritt}, D., {Wandel},
  A., \& {Joseph}, C.~L. 2001, \apjl, 555, L79

\bibitem[{{Figer} {et~al.}(1999{\natexlab{a}}){Figer}, {Kim}, {Morris},
  {Serabyn}, {Rich}, \& {McLean}}]{FKMSRM99}
{Figer}, D.~F., {Kim}, S.~S., {Morris}, M., {Serabyn}, E., {Rich}, R.~M., \&
  {McLean}, I.~S. 1999{\natexlab{a}}, \apj, 525, 750

\bibitem[{{Figer} {et~al.}(1999{\natexlab{b}}){Figer}, {McLean}, \&
  {Morris}}]{FMcLM99}
{Figer}, D.~F., {McLean}, I.~S., \& {Morris}. 1999{\natexlab{b}}, \apj, 514,
  202

\bibitem[{{Figer} {et~al.}(1998){Figer}, {Najarro}, {Morris}, {McLean},
  {Geballe}, {Ghez}, \& {Langer}}]{FigerEtAl98}
{Figer}, D.~F., {Najarro}, F., {Morris}, M., {McLean}, I.~S., {Geballe}, T.~R.,
  {Ghez}, A.~M., \& {Langer}, N. 1998, \apj, 506, 384

\bibitem[{{Fregeau} {et~al.}(2003{\natexlab{a}}){Fregeau}, {G\"urkan}, {Joshi},
  \& {Rasio}}]{FGJR03}
{Fregeau}, J.~M., {G\"urkan}, M.~A., {Joshi}, K.~J., \& {Rasio}, F.~A.
  2003, \apj, 593, 772

\bibitem[{{Fregeau} {et~al.}(2003{\natexlab{b}}){Fregeau}, {G\"urkan}, \&
  {Rasio}}]{FGR03b}
{Fregeau}, J.~M., {G\"urkan}, M.~A., \& {Rasio}, F.~A. 2004,
  {Formation of Massive Black Holes in Dense Star Clusters. III. Effects of
  Primordial Binary Stars.}, in preparation

\bibitem[{{Fregeau} {et~al.}(2002){Fregeau}, {Joshi}, {Portegies Zwart}, \&
  {Rasio}}]{FJPZR02}
{Fregeau}, J.~M., {Joshi}, K.~J., {Portegies Zwart}, S.~F., \& {Rasio}, F.~A.
  2002, \apj, 570, 171

\bibitem[{{Freitag} {et~al.}(2003{\natexlab{a}}){Freitag}, {Amaro-Seoane}, \&
  {Spurzem}}]{FASS03}
{Freitag}, M., {Amaro-Seoane}, P., \& {Spurzem}, R. 2003{\natexlab{a}}, {The
  mechanism of core collapse in stellar clusters with a broad mass spectrum},
  in preparation

\bibitem[{{Freitag} \& {Benz}(2001)}]{FB01}
{Freitag}, M., \& {Benz}, W. 2001, \aap, 375, 711

\bibitem[{{Freitag} \& {Benz}(2002)}]{FB02}
---. 2002, \aap, 394, 345

\bibitem[{{Freitag} {et~al.}(2003{\natexlab{b}}){Freitag}, {G\"urkan}, \&
  {Rasio}}]{FGR03}
{Freitag}, M., {G\"urkan}, M.~A., \& {Rasio}, F.~A. 2003{\natexlab{b}},
  {Formation of massive black holes in dense star clusters II: Collisional
  run-away in proto-galactic nuclei}, in preparation

\bibitem[{{Gallagher} \& {Smith}(1999)}]{GS99}
{Gallagher}, J.~S., \& {Smith}, L.~J. 1999, \mnras, 304, 540

%
\bibitem[{{Gebhardt} {et~al.}(2000)}]{GebhardtEtAl00}
{Gebhardt}, K., et al. 2000, \apjl, 539, L13

\bibitem[{{Gebhardt} {et~al.}(2002){Gebhardt}, {Rich}, \& {Ho}}]{GRH02}
{Gebhardt}, K., {Rich}, R.~M., \& {Ho}, L.~C. 2002, \apjl, 578, L41

\bibitem[{{Gerssen} {et~al.}(2002){Gerssen}, {van der Marel}, {Gebhardt},
  {Guhathakurta}, {Peterson}, \& {Pryor}}]{Getal02}
{Gerssen}, J., {van der Marel}, R.~P., {Gebhardt}, K., {Guhathakurta}, P.,
  {Peterson}, R.~C., \& {Pryor}, C. 2002, \aj, 124, 3270

\bibitem[{{Gerssen} {et~al.}(2003){Gerssen}, {van der Marel}, {Gebhardt},
  {Guhathakurta}, {Peterson}, \& {Pryor}}]{Getal03}
---. 2003, \aj, 125, 376

\bibitem[{{Ghez} {et~al.}(2000){Ghez}, {Morris}, {Becklin}, {Tanner}, \&
  {Kremenek}}]{GhezEtAl00}
{Ghez}, A.~M., {Morris}, M., {Becklin}, E.~E., {Tanner}, A., \& {Kremenek}, T.
  2000, \nat, 407, 349

\bibitem[{{Ghez} {et~al.}(2003){Ghez}, {Salim}, {Hornstein}, {Tanner}, {ris},
  {Becklin}, \& {Duchene}}]{GhezEtAl03}
{Ghez}, A.~M., {Salim}, S., {Hornstein}, S.~D., {Tanner}, A., {ris}, M.~M.,
  {Becklin}, E.~E., \& {Duchene}, G. 2003, {Stellar Orbits Around the Galactic
  Center Black Hole}, eprint astro-ph/0306130

\bibitem[{{Giersz}(1998)}]{Giersz98}
{Giersz}, M. 1998, \mnras, 298, 1239

\bibitem[{{Giersz}(2001)}]{Giersz01}
---. 2001, \mnras, 324, 218

\bibitem[{{Giersz} \& {Heggie}(1996)}]{GH96}
{Giersz}, M., \& {Heggie}, D.~C. 1996, \mnras, 279, 1037

\bibitem[{{Giersz} \& {Heggie}(1997)}]{GH97}
---. 1997, \mnras, 286, 709

\bibitem[{{Giersz} \& {Spurzem}(1994)}]{GS94}
{Giersz}, M., \& {Spurzem}, R. 1994, \mnras, 269, 241

\bibitem[{{Giersz} \& {Spurzem}(2003)}]{GS03}
---. 2003, {A stochastic Monte Carlo approach to model real star cluster
  evolution, III. Direct integrations of three- and four-body interactions},
  submitted to \mnras, astro-ph/0301643

\bibitem[{{G\"urkan} \& {Rasio}(2003)}]{GR03}
{G\"urkan}, M.~A., \& {Rasio}, F.~A. 2003, {Intermediate-Mass Black Holes as
  Remnants of Disrupted Star Clusters}, in preparation

\bibitem[{{Hansen} \& {Milosavljevi{\'c}}(2003)}]{HM03}
{Hansen}, B., \& {Milosavljevi{\'c}}, M. 2003, \apjl,  593, L77

\bibitem[{{Heger} {et~al.}(2002){Heger}, {Fryer}, {Woosley}, {Langer}, \&
  {Hartmann}}]{HFWLH02}
{Heger}, A., {Fryer}, C.~L., {Woosley}, S.~E., {Langer}, N., \& {Hartmann},
  D.~H. 2002, {How Massive Single Stars End their Life}, submitted to ApJ,
  astro-ph/0212469

\bibitem[{{Heger} {et~al.}(2003){Heger}, {Woosley}, {Fryer}, \&
  {Langer}}]{HWFL03}
{Heger}, A., {Woosley}, S.~E., {Fryer}, C.~L., \& {Langer}, N. 2003, in From
  Twilight to Highlight: The Physics of Supernovae. Proceedings of the
  ESO/MPA/MPE Workshop held in Garching, Germany, 3, astro-ph/0211062

\bibitem[{{Heggie} \& {Hut}(2003)}]{HH03}
{Heggie}, D., \& {Hut}, P. 2003, {The Gravitational Million-Body Problem: A
  Multidisciplinary Approach to Star Cluster Dynamics} (Cambridge University
  Press)

\bibitem[{{Heggie}(1975)}]{Heggie75}
{Heggie}, D.~C. 1975, \mnras, 173, 729

\bibitem[{{H{\' e}non}(1971{\natexlab{a}})}]{Henon71}
{H{\' e}non}, M.~H. 1971{\natexlab{a}}, \apss, 13, 284

\bibitem[{{H{\' e}non}(1971{\natexlab{b}})}]{Henon71b}
---. 1971{\natexlab{b}}, \apss, 14, 151

\bibitem[{{H{\' e}non}(1973)}]{Henon73}
{H{\' e}non}, M.~H. 1973, in Saas-Fee Advanced Course 3: Dynamical Structure
  and Evolution of Stellar Systems, 183

\bibitem[{{Ho}(1998)}]{Ho98}
{Ho}, L.~C. 1998, in Observational Evidence for Black Holes in the Universe,
  ed. S. K. Chakrabarti (Dordrecht: Kluwer), astro-ph/9803307

\bibitem[{{Ho} \& {Filippenko}(1996)}]{HF96}
{Ho}, L.~C., \& {Filippenko}, A.~V. 1996, \apjl, 466, L83

\bibitem[{{Hurley} {et~al.}(2000){Hurley}, {Pols}, \& {Tout}}]{HPT00}
{Hurley}, J.~R., {Pols}, O.~R., \& {Tout}, C.~A. 2000, \mnras, 315, 543

\bibitem[{{Hut}(1985)}]{Hut85}
{Hut}, P. 1985, in IAU Symp. 113: Dynamics of Star Clusters, 231--247

\bibitem[{{Inagaki}(1985)}]{Inagaki85}
{Inagaki}, S. 1985, in IAU Symp. 113: Dynamics of Star Clusters, 189--204

\bibitem[{{Inagaki} \& {Saslaw}(1985)}]{IS85}
{Inagaki}, S., \& {Saslaw}, W.~C. 1985, \apj, 292, 339

\bibitem[{{Inagaki} \& {Wiyanto}(1984)}]{IW84}
{Inagaki}, S., \& {Wiyanto}, P. 1984, \pasj, 36, 391

\bibitem[{{Joshi} {et~al.}(2001){Joshi}, {Nave}, \& {Rasio}}]{JNR01}
{Joshi}, K.~J., {Nave}, C.~P., \& {Rasio}, F.~A. 2001, \apj, 550, 691

\bibitem[{{Joshi} {et~al.}(2000){Joshi}, {Rasio}, \& {Portegies
  Zwart}}]{JRPZ00}
{Joshi}, K.~J., {Rasio}, F.~A., \& {Portegies Zwart}, S. 2000, \apj, 540, 969

\bibitem[{{Kaaret} {et~al.}(2001){Kaaret}, {Prestwich}, {Zezas}, {Murray},
  {Kim}, {Kilgard}, {Schlegel}, \& {Ward}}]{Kaaretetal01}
{Kaaret}, P., {Prestwich}, A.~H., {Zezas}, A., {Murray}, S.~S., {Kim}, D.-W.,
  {Kilgard}, R.~E., {Schlegel}, E.~M., \& {Ward}, M.~J. 2001, \mnras, 321, L29

\bibitem[{{Kim} {et~al.}(2002){Kim}, {Einsel}, {Lee}, {Spurzem}, \&
  {Lee}}]{KELSL02}
{Kim}, E., {Einsel}, C., {Lee}, H.~M., {Spurzem}, R., \& {Lee}, M.~G. 2002,
  \mnras, 334, 310

\bibitem[{{King} {et~al.}(2001){King}, {Davies}, {Ward}, {Fabbiano}, \&
  {Elvis}}]{Kingetal01}
{King}, A.~R., {Davies}, M.~B., {Ward}, M.~J., {Fabbiano}, G., \& {Elvis}, M.
  2001, \apjl, 552, L109

\bibitem[{{Kormendy} \& {Gebhardt}(2001)}]{KG01}
{Kormendy}, J., \& {Gebhardt}, K. 2001, in 20th Texas Symposium on relativistic
  astrophysics, ed. H.~{Martel} \& J.~C. {Wheeler}, 363

\bibitem[{{Kroupa}(2002)}]{Kroupa02}
{Kroupa}, P. 2002, Science, 295, 82

\bibitem[{{Kroupa} {et~al.}(1993){Kroupa}, {Tout}, \& {Gilmore}}]{KTG93}
{Kroupa}, P., {Tout}, C.~A., \& {Gilmore}, G. 1993, \mnras, 262, 545

\bibitem[{{Kulkarni} {et~al.}(1993){Kulkarni}, {Hut}, \& {McMillan}}]{KHMcM93}
{Kulkarni}, S.~R., {Hut}, P., \& {McMillan}, S. 1993, \nat, 364, 421

\bibitem[{{Larson}(1970)}]{Larson70}
{Larson}, R.~B. 1970, \mnras, 150, 93

\bibitem[{{Lauer} {et~al.}(1998){Lauer}, {Faber}, {Ajhar}, {Grillmair}, \&
  {Scowen}}]{LFAGS98}
{Lauer}, T.~R., {Faber}, S.~M., {Ajhar}, E.~A., {Grillmair}, C.~J., \&
  {Scowen}, P.~A. 1998, \aj, 116, 2263

\bibitem[{{Lee}(1995)}]{Lee95}
{Lee}, H.~M. 1995, \mnras, 272, 605

\bibitem[{{Lee}(2001)}]{Lee01}
---. 2001, Classical and Quantum Gravity, 18, 3977

\bibitem[{{Lightman} \& {McMillan}(1985)}]{LM85}
{Lightman}, A.~P., \& {McMillan}, S.~L.~W. 1985, in IAU Symp. 113: Dynamics of
  Star Clusters, 261--274

\bibitem[{{Louis} \& {Spurzem}(1991)}]{LS91}
{Louis}, P.~D., \& {Spurzem}, R. 1991, \mnras, 251, 408

\bibitem[{{Lynden-Bell} \& {Wood}(1968)}]{LBW68}
{Lynden-Bell}, D., \& {Wood}, R. 1968, \mnras, 138, 495

\bibitem[{{Maeder} \& {Meynet}(2001)}]{MM01}
{Maeder}, A., \& {Meynet}, G. 2001, \aap, 373, 555

\bibitem[{{Makino}(2001)}]{Makino01}
{Makino}, J. 2001, in ASP Conf. Ser. 228: Dynamics of Star Clusters and the
  Milky Way, 87

\bibitem[{{Makino}(2002)}]{Makino02}
{Makino}, J. 2002, in ASP Conf. Ser. 263: Stellar Collisions, Mergers and their
  Consequences, 389

\bibitem[{{Malyshkin} \& {Goodman}(2001)}]{MG02}
{Malyshkin}, L., \& {Goodman}, J. 2001, Icarus, 150, 314

\bibitem[{{Meylan} {et~al.}(2001){Meylan}, {Sarajedini}, {Jablonka},
  {Djorgovski}, {Bridges}, \& {Rich}}]{MeylanEtAl01}
{Meylan}, G., {Sarajedini}, A., {Jablonka}, P., {Djorgovski}, S.~G., {Bridges},
  T., \& {Rich}, R.~M. 2001, \aj, 122, 830

\bibitem[{{Meynet} \& {Maeder}(2000)}]{MM00}
{Meynet}, G., \& {Maeder}, A. 2000, \aap, 361, 101

\bibitem[{{Miller} \& {Scalo}(1979)}]{MS79}
{Miller}, G.~E., \& {Scalo}, J.~M. 1979, \apjs, 41, 513

\bibitem[{{Miller} {et~al.}(2003){Miller}, {Fabbiano}, {Miller}, \&
  {Fabian}}]{MFMF03}
{Miller}, J.~M., {Fabbiano}, G., {Miller}, M.~C., \& {Fabian}, A.~C. 2003,
  \apjl, 585, L37

\bibitem[{{Miller} \& {Hamilton}(2002)}]{MH02}
{Miller}, M.~C., \& {Hamilton}, D.~P. 2002, \mnras, 330, 232

\bibitem[{{Murray} \& {Lin}(1996)}]{ML96}
{Murray}, S.~D., \& {Lin}, D.~N.~C. 1996, \apj, 467, 728

\bibitem[{{Nakamura} \& {Umemura}(2001)}]{NU01}
{Nakamura}, F., \& {Umemura}, M. 2001, \apj, 548, 19

\bibitem[{{Portegies Zwart} {et~al.}(1999){Portegies Zwart}, {Makino},
  {McMillan}, \& {Hut}}]{PZMMH99}
{Portegies Zwart}, S.~F., {Makino}, J., {McMillan}, S. L.~W., \& {Hut}, P.
  1999, \aa, 348, 117

\bibitem[{{Portegies Zwart} \& {McMillan}(2000)}]{PZMcM00}
{Portegies Zwart}, S.~F., \& {McMillan}, S.~L.~W. 2000, \apjl, 528, 17

\bibitem[{{Portegies Zwart} \& {McMillan}(2002)}]{PZM02}
---. 2002, \apj, 576, 899

\bibitem[{{Quinlan}(1996)}]{Quinlan96}
{Quinlan}, G.~D. 1996, New Astronomy, 1, 255

\bibitem[{{Quinlan} \& {Shapiro}(1989)}]{QS89}
{Quinlan}, G.~D., \& {Shapiro}, S.~L. 1989, \apj, 343, 725

\bibitem[{{Raboud} \& {Mermilliod}(1998)}]{RM98}
{Raboud}, D., \& {Mermilliod}, J.-C. 1998, \aap, 333, 897

\bibitem[{{Rasio} {et~al.}(2003){Rasio}, {Freitag}, \& {G\"urkan}}]{RFG03}
{Rasio}, F.~A., {Freitag}, M., \& {G\"urkan}, M.~A. 2003, in Carnegie
  Observatories Astrophysics Series, Vol. 1: Coevolution of Black Holes and
  Galaxies, astro-ph/0304038

\bibitem[{{Rees}(1984)}]{Rees84}
{Rees}, M.~J. 1984, \araa, 22, 471

\bibitem[{{Saslaw} \& {De Young}(1971)}]{SDY71}
{Saslaw}, W.~C., \& {De Young}, D.~S. 1971, \apj, 170, 423

%
\bibitem[{{Sch{\" o}del} {et~al.}(2002)}]{SchodelEtAl02}
{Sch{\" o}del}, R., et al.  2002, \nat, 419, 694

\bibitem[{{Schaller} {et~al.}(1992){Schaller}, {Schaerer}, {Meynet}, \&
  {Maeder}}]{SSMM92}
{Schaller}, G., {Schaerer}, D., {Meynet}, G., \& {Maeder}, A. 1992, \aap, 96,
  269

\bibitem[{{Sigurdsson} \& {Hernquist}(1993)}]{SH93}
{Sigurdsson}, S., \& {Hernquist}, L. 1993, \nat, 364, 423

\bibitem[{{Spinnato} {et~al.}(2002){Spinnato}, {Fellhauer}, \&
  Portegies~{Zwart}}]{SFPZ02}
{Spinnato}, P.~F., {Fellhauer}, M., \& Portegies~{Zwart}, S.~F. 2002, {The
  efficiency of the spiral-in of a black hole to the Galactic centre}, eprint
  astro-ph/0202494

\bibitem[{{Spitzer}(1987)}]{Spitzer87}
{Spitzer}, L. 1987, {Dynamical evolution of globular clusters} (Princeton, NJ,
  Princeton University Press)

\bibitem[{{Spitzer} \& {Shull}(1975)}]{SS75}
{Spitzer}, L., \& {Shull}, J.~M. 1975, \apj, 201, 773

\bibitem[{{Spitzer}(1969)}]{Spitzer69}
{Spitzer}, L.~J. 1969, \apjl, 158, L139

\bibitem[{{Spurzem} \& {Baumgardt}(2003)}]{SB03}
{Spurzem}, R., \& {Baumgardt}, H. 2003, {A parallel implementation of a direct
  {$N$}-body integrator on general and special purpose supercomputers}, in
  preparation

\bibitem[{{Spurzem} \& {Takahashi}(1995)}]{ST95}
{Spurzem}, R., \& {Takahashi}, K. 1995, \mnras, 272, 772

\bibitem[{{Stod{\'o}{\l}kiewicz}(1982)}]{Stodol82}
{Stod{\'o}{\l}kiewicz}, J.~S. 1982, Acta Astronomica, 32, 63

\bibitem[{{Stod{\'o}{\l}kiewicz}(1986)}]{Stodol86}
---. 1986, Acta Astronomica, 36, 19

\bibitem[{{Takahashi}(1997)}]{Takahashi97}
{Takahashi}, K. 1997, \pasj, 49, 547

%
\bibitem[{{Tremaine} {et~al.}(2002)}]{Tetal02}
{Tremaine}, S., et al. 2002, \apj, 574, 740

\bibitem[{{Tremaine} {et~al.}(1994){Tremaine}, {Richstone}, {Byun}, {Dressler},
  {Faber}, {Grillmair}, {Kormendy}, \& {Lauer}}]{Tetal94}
{Tremaine}, S., {Richstone}, D.~O., {Byun}, Y., {Dressler}, A., {Faber}, S.~M.,
  {Grillmair}, C., {Kormendy}, J., \& {Lauer}, T.~R. 1994, \aj, 107, 634

\bibitem[{{van der Marel}(2001)}]{vdM01}
{van der Marel}, R.~P. 2001, in Black Holes in Binaries and Galactic Nuclei,
  246

\bibitem[{{van der Marel}(2003)}]{vdM03}
{van der Marel}, R.~P. 2003, in Carnegie Observatories Astrophysics Series,
  Vol. 1: Coevolution of Black Holes and Galaxies, astro-ph/0302101

\bibitem[{{Vishniac}(1978)}]{Vishniac78}
{Vishniac}, E.~T. 1978, \apj, 223, 986

\bibitem[{{Watters} {et~al.}(2000){Watters}, {Joshi}, \& {Rasio}}]{WJR00}
{Watters}, W.~A., {Joshi}, K.~J., \& {Rasio}, F.~A. 2000, \apj, 539, 331

\bibitem[{{Wielen}(1975)}]{Wielen75}
{Wielen}, R. 1975, in IAU Symp. 69: Dynamics of the Solar Systems, 119--131

\bibitem[{{Zezas} \& {Fabbiano}(2002)}]{ZF02}
{Zezas}, A., \& {Fabbiano}, G. 2002, \apj, 577, 726

\end{thebibliography}

\end{document}